# Thermal fluctuation field for current-induced domain wall motion

Kyoung-Whan Kim and Hyun-Woo Lee

*PCTP and Department of Physics, Pohang University of Science and Technology, Pohang 790-784, Korea*

Current-induced domain wall motion in magnetic nanowires is affected by thermal fluctuation. In order to account for this effect, the Landau-Lifshitz-Gilbert equation includes a thermal fluctuation field and literature often utilizes the fluctuation-dissipation theorem to characterize statistical properties of the thermal fluctuation field. However, the theorem is not applicable to the system under finite current since it is not in equilibrium. To examine the effect of finite current on the thermal fluctuation, we adopt the influence functional formalism developed by Feynman and Vernon, which is known to be a useful tool to analyze effects of dissipation and thermal fluctuation. For this purpose, we construct a quantum-mechanical effective Hamiltonian describing current-induced domain wall motion by generalizing the Caldeira-Leggett description of quantum dissipation. We find that even for the current-induced domain wall motion, the statistical properties of the thermal noise is still described by the fluctuation-dissipation theorem if the current density is sufficiently lower than the intrinsic critical current density and thus the domain wall tilting angle is sufficiently lower than $\pi/4$. The relation between our result and a recent result [R. A. Duine, A. S. Núñez, J. Sinova, and A. H. MacDonald, Phys. Rev. B **75**, 214420 (2007)], which also addresses the thermal fluctuation, is discussed. We also find interesting physical meanings of the Gilbert damping $\alpha$ and the nonadiabaticy parameter $\beta$; while $\alpha$ characterizes the coupling strength between the magnetization dynamics (the domain wall motion in this paper) and the thermal reservoir (or environment), $\beta$ characterizes the coupling strength between the spin current and the thermal reservoir.

## I. INTRODUCTION

Current-induced domain wall (DW) motion in a ferromagnetic nanowire is one of representative examples to study the effect of spin-transfer torque (STT). The motion of DW is generated by the angular momentum transfer between space-time-dependent magnetization $\vec{m}(x,t)$ and conduction electrons, of which spins interact with $\vec{m}$ by the exchange coupling. This system is usually described by the Landau-Lifshitz-Gilbert (LLG) equation,[1–3]

$$\frac{\partial \vec{m}}{\partial t} = \gamma_0 \vec{H}_{eff} \times \vec{m} + \frac{\alpha}{m_s} \vec{m} \times \frac{\partial \vec{m}}{\partial t} + \frac{j_p \mu_B}{e m_s} \left[ \frac{\partial \vec{m}}{\partial x} - \frac{\beta}{m_s} \vec{m} \times \frac{\partial \vec{m}}{\partial x} \right], \quad (1)$$

where $\gamma_0$ is the gyromagnetic ratio, $j_p$ is the spin-current density, $m_s = |\vec{m}|$ is the saturation magnetization, and $\mu_B$ is the Bohr magneton. $\alpha$ is the Gilbert damping coefficient, and $\beta$ is the nonadiabatic coefficient representing the magnitude of the nonadiabatic STT.[4] In Eq. (1), the effective magnetic field $H_{eff}$ is given by

$$\vec{H}_{eff} = A \nabla^2 \vec{m} + \vec{H}_{ani} + \vec{H}_{th}, \quad (2)$$

where $A$ is stiffness constant, $\vec{H}_{ani}$ describes the effect of the magnetic anisotropy, and $\vec{H}_{th}$ is the thermal fluctuation field describing the thermal noise. In equilibrium situations, the magnitude and spatiotemporal correlation of $\vec{H}_{th}$ are governed by the fluctuation-dissipation theorem,[5–7]

$$\langle H_{th,i}(\vec{x},t) H_{th,j}(\vec{x}',t') \rangle = \frac{4\alpha k_B T}{\hbar \rho} \delta(\vec{x}-\vec{x}') \delta(t-t') \delta_{ij}, \quad (3)$$

where $\langle \cdots \rangle$ represents the statistical average, $i,j$ denote $x$, $y$, or $z$ component, $k_B$ is the Boltzmann constant, $T$ is the temperature, and $\rho = m_s/\mu_B$ is the spin density. Equation (3) plays an important role for the study of the magnetization dynamics at finite temperature.[8]

Equation (3) has been also used in literature[9–13] to examine effects of thermal fluctuations on the current-induced DW motion. In nonequilibrium situations, however, the fluctuation-dissipation theorem does not hold generally. Since the system is not in equilibrium any more when the current is applied, it is not clear whether Eq. (3) may be still used. Recalling that $\vec{H}_{th}$ is estimated to affect the magnetization dynamics considerably in many experimental situations[14–17] of the current-driven DW motion, it is highly desired to properly characterize $\vec{H}_{th}$ in situations with nonzero $j_p$. Recently, Duine[18] attempted this characterization and showed that Eq. (3) is not altered by the spin current up to first order in the spin-current magnitude. This analysis however is limited to situations where the spin-flip scattering is the main mechanism responsible for $\beta$. In this paper, we generalize this analysis by using a completely different approach which does not assume any specific physical origin of $\beta$.

$H_{th}$ arises from extra degrees of freedom (other than magnetization), which are not included in the LLG equation. The extra degrees of freedom (phonons for instance) usually have much larger number of degrees of freedom than magnetization and thus form a heat reservoir. Thus properties of $H_{th}$ are determined by the heat reservoir. The heat reservoir plays another role. In the absence of the extra degrees of freedom, the Gilbert damping coefficient $\alpha$ should be zero since the total energy should be conserved when all degrees of freedom are taken into account. Thus the heat reservoir is responsible also for finite $\alpha$. These dual roles of the heat reservoir are the main idea behind the Einstein's theory of the Brownian motion.[19] There are also claims that $\alpha$ is correlated with $\beta$ (Refs. [18] and [20–22]) in the sense that mechanisms, which generate $\beta$, also contribute to $\alpha$. Thus the issue of $\vec{H}_{th}$ and the issue of $\alpha$ and $\beta$ are mutually connected. Recalling that the main mechanism responsible for $\alpha$ varies from material to material, it is reasonable to expect that the main mechanism for $\vec{H}_{th}$ and $\beta$ may also vary from material to material. Recently, various mechanisms of $\beta$ were examined such as momentum transfer,[23–25] spin mistracking,[26,27] spin-flip scattering,[18,21,22,25,28] and the influence of a transport current.[29] This diversity of mechanisms will probably apply to $\vec{H}_{th}$ as well.

Instead of examining each mechanism of $\vec{H}_{th}$ one by one, we take an alternative approach to address this issue. In 1963, Feynman and Vernon[30] proposed the so-called influence functional formalism, which allows one to take account of damping effects without detailed accounts of damping mechanisms. This formalism was later generalized by Smith and Caldeira.[31] This formalism has been demonstrated to be a useful tool to address dissipation effects (without specific accounts of detailed damping mechanisms) on, for instance, quantum tunneling,[32] nonequilibrium dynamic Coulomb blockade,[33] and quantum noise.[34] To take account of damping effects which are energy nonconserving processes in general, the basic idea of the influence functional formalism is to introduce infinite number of degrees of freedom (called *environment*) behaves like harmonic oscillators which couple with the damped system. [See Eq. (13).] Caldeira and Leggett[32] suggested the structure of the spectrum of environment Eq. (13) and integrated out the degrees of freedom of environment to find the effective Hamiltonian describing the classical damping Eq. (12). For readers who are not familiar with the Caldeira-Leggett's theory of quantum dissipation, we present the summary of details of the theory in Sec. II B.

In order to address the issue of $\vec{H}_{th}$, we follow the idea of the influence functional formalism and construct an effective Hamiltonian describing the magnetization dynamics. The effective Hamiltonian describes not only energy-conserving processes but also energy-nonconserving processes such as damping and STT. From this approach, we find that Eq. (3) holds even in nonequilibrium situations with finite $j_p$, provided that $j_p$ is sufficiently smaller than the so-called intrinsic critical current density[23] so that the DW tilting angle $\phi$ (to be defined below) is sufficiently smaller than $\pi/4$. We remark that in the special case where the spin-flip scattering mechanism of $\beta$ is the main mechanism of $\vec{H}_{th}$, our finding is consistent with Ref. [18], which reports that the spin flip scattering mechanism does not alter Eq. (3) at least up to the first order in $j_p$. But our calculation indicates that Eq. (3) holds not only in situations where the spin flip scattering is the dominant mechanism of $\vec{H}_{th}$ and $\beta$ but also in more diverse situations as long as the heat reservoir can be described by bosonic excitations (such as electron-hole pair excitations or phonon), i.e., the excitations effectively behave like harmonic oscillators to be described by Caldeira-Leggett's theory. We also remark that in addition to the derivation of Eq. (3) in nonequilibrium situations, our calculation also reveals an interesting physical meaning of $\beta$, which will be detailed in Sec. III.

This paper is organized as follows. In Sec. II, we first introduce the Caldeira-Leggett's version of the influence functional formalism and later generalize this formalism so that it is applicable to our problem. This way, we construct a Hamiltonian describing the DW motion. In Sec. III, some implications of this model is discussed. First, a distinct insight on $\beta$ is emphasized. Second, as an application, statistical properties of the thermal fluctuation field are calculated in the presence of nonzero $j_p$, which verifies the validity of Eq. (3) when $j_p$ is sufficiently smaller than the intrinsic critical density. It is believed that many experiments[16,17] are indeed in this regime. Finally, in Sec. IV, we present some concluding remarks. Technical details about the quantum theory of the DW motion and methods to obtain solutions are included in Appendices.

## II. GENERALIZED CALDEIRA-LEGGETT DESCRIPTION

### A. Background

Instead of full magnetization profile $\vec{m}(x,t)$, the DW dynamics is often described[2,23,35–37] by two collective coordinates, DW position $x(t)$ and DW tilting angle $\phi(t)$. When expressed in terms of these collective coordinates, the LLG Eq. (1) reduces to the so-called Thiele equations,

$$\frac{dx}{dt} = \frac{j_p \mu_B}{e m_s} + \alpha\lambda\frac{d\phi}{dt} + \frac{\gamma_0 K \lambda}{m_s}\sin 2\phi + \eta_x(t), \quad (4a)$$

$$\lambda\frac{d\phi}{dt} = -\alpha\frac{dx}{dt} + \beta\frac{j_p \mu_B}{e m_s} + \eta_p(t). \quad (4b)$$

Here $K$ is the hard-axis anisotropy, $\lambda$ is the DW thickness. $\eta_x(t)$ and $\eta_p(t)$ are functions describing thermal noise field $H_{th,i}(x,t)$. By definition, the statistical average of the thermal noise field $H_{th,i}(x,t)$ is zero and similarly the statistical averages of $\eta_x(t)$ and $\eta_p(t)$ should also vanish regardless of whether the system is in equilibrium. The question of their correlation function is not trivial however. If the thermal noise field $H_{th,i}(x,t)$ satisfies the correlation in Eq. (3), it can be derived from Eq. (3) that $\eta_x(t)$ and $\eta_p(t)$ satisfy the correlation relation[12]

$$\langle \eta_i(t)\eta_j(t')\rangle \propto \alpha k_B T \delta_{ij}\delta(t-t'), \quad (5)$$

for $\{i,j\}=\{x,p\}$. But as mentioned in Sec. I, Eq. (3) is not guaranteed generally in the presence of the nonzero current. Then Eq. (5) is not guaranteed either. The question of what should be the correlation function $\langle \eta_i(t)\eta_j(t')\rangle$ in such a situation will be discussed in Sec. III.

When the spin-current density $j_p$ is sufficiently smaller than the so-called intrinsic critical density $|e\gamma_0 K\lambda/\mu_B|$,[23] $\phi$ stays sufficiently smaller than $\pi/4$. In many experimental situations,[38–40] this is indeed the case,[41] so we will confine ourselves to the small $\phi$ regime in this paper. Then, one can approximate $\sin 2\phi \approx 2\phi$ to convert the equations into the following form:[42]

$$\frac{dx}{dt} = v_s + \frac{\alpha S}{2KM}\frac{dp}{dt} + \frac{p}{M} + \eta_1(t), \quad (6a)$$

$$\frac{dp}{dt} = -\frac{2\alpha KM}{S}\frac{dx}{dt} + \frac{2\beta KM}{S}v_s + \eta_2(t), \quad (6b)$$

where $p = 2KM\lambda\phi/S$, $S$ is the spin angular momentum at each individual magnetic site, and $v_s = j_p \mu_B / e m_s$ is the adiabatic velocity,[43] which is a constant of velocity dimension and proportional to $j_p$. The yet undetermined constant $M$ is the effective DW mass[42–44] which will be fixed so that the new variable $p$ becomes the canonical conjugate to $x$. $\eta_1(t)$ and $\eta_2(t)$ are the same as $\eta_x(t)$ and $\eta_p(t)$ except for proportionality constants.

When the thermal noises $\eta_1(t)$ and $\eta_2(t)$ are ignored, one obtains from Eq. (6) the time dependence of the DW position,

$$x(t) = x(0) + \frac{\beta}{\alpha}v_s t + \frac{S}{2KM\alpha^2}(1 - e^{-2K\alpha t/S(1+\alpha^2)}) \\ \times [\alpha p(0) - Mv_s(\alpha - \beta)]. \quad (7)$$

Note that after a short transient time, the DW speed approaches the terminal velocity $\beta v_s/\alpha$. Thus the ratio $\beta/\alpha$ is an important parameter for the DW motion. When the thermal noises are considered, they generate a correction to Eq. (7). However, from Eq. (6), it is evident that the statistical average of $x(t)$ should still follow Eq. (7). Thus as far as the temporal evolution of the statistical average is concerned, we may ignore the thermal noises. In the rest of Sec. II, we aim to derive a quantum mechanical Hamiltonian, which reproduces the same temporal evolution as Eq. (7) in the statistical average level. In Sec. III, we use the Hamiltonian to derive the correlation function $\langle \eta_i(t)\eta_j(t')\rangle$ in the presence of the nonzero current.

Now, we begin our attempt to construct an effective Hamiltonian that reproduces the DW dynamics Eq. (6) [or equivalently Eq. (7)]. We first begin with the microscopic quantum-mechanical Hamiltonian $H_{s-d}$,

$$H_{s-d} = -J\sum_i \vec{S}_i \cdot \vec{S}_{i+1} - A\sum_i (\vec{S}_i \cdot \hat{z})^2 + K\sum_i (\vec{S}_i \cdot \hat{y})^2 + H_{cS}, \quad (8)$$

which has been used in previous studies[20] of the DW dynamics. Here $J$ represents the ferromagnetic exchange interaction, $A$ and $K$ represent longitudinal (easy-axis) and transverse (hard-axis) anisotropy, respectively. The last term $H_{cS}$ represents the coupling of the spin system with the spin-polarized current,

$$H_{cS} = -\sum_{i,\alpha=\uparrow,\downarrow}[t(c^\dagger_{i\alpha}c_{i+1\alpha} + c^\dagger_{i+1\alpha}c_{i\alpha}) - \mu c^\dagger_{i\alpha}c_{i\alpha}] - J_H\sum_i \vec{S}_{ci} \cdot \vec{S}_i, \quad (9)$$

where $J_H$ is the exchange interaction between conduction electron and the localized spins, $c_{i\alpha}$ is the annihilation operator of the conduction electron at the site $i$, $\vec{S}_{ci}$ is the electron-spin operator, $t$ is the hopping integral, and $\mu$ is the chemical potential of the system.

Recently Kim et al.[43] analyzed $H_{s-d}$ in detail in the small tilting angle regime and found that $H_{s-d}$ contains gapless low-lying excitations and also high-energy excitations with a finite energy gap. The gapless excitations of $H_{s-d}$ are described by a simple Hamiltonian $H_0$,

$$H_0 = v_s P + \frac{P^2}{2M} \quad (10)$$

while the high-energy excitations have a finite energy gap $2S\sqrt{A(A+K)}$. In Eq. (10), $P$ is the canonical momentum of the DW position operator $Q$, and $M = \frac{\hbar^2}{K}\sqrt{\frac{2A}{Ja^4}}$ is the effective DW mass called Döring mass.[44] Here, $a$ is the lattice spacing between two neighboring spins. (See, for details, Appendix A.) Below we will neglect the high energy excitations and focus on the low-lying excitations described by Eq. (10). For the analysis of the high-energy excitation effects on the DW, See Ref. 42.

From Eq. (10), one obtains the following Heisenberg's equation of motion:

$$\frac{dQ}{dt} = v_s + \frac{P}{M}, \quad (11a)$$

$$\frac{dP}{dt} = 0. \quad (11b)$$

Note that the current (proportional to $v_s$) appears in the equation for $\frac{dQ}{dt}$. Thus the current affects the DW dynamics by introducing a difference between the canonical momentum $P$ and the kinematic momentum $P + Mv_s$. In this sense, the effect of the current is similar to a vector potential [canonical momentum $\vec{P}$ and kinematic momentum $\vec{P} + (e/c)\vec{A}$]. The vector potential (difference between the canonical momentum and the kinetic momentum) allows the system in the initially zero momentum state to move without breaking the translational symmetry of the system. In other words, the current-induced DW motion is generated without any force term in Eq. (11b) violating the translational symmetry of the system. This should be contrasted with the effect of the magnetic field or magnetic defects, which generates a force term in Eq. (11b).

The solution of Eq. (11) is trivial, $\langle Q(t)\rangle = \langle Q(0)\rangle + (\langle P(0)\rangle/M + v_s)t$. Here, the statistical average $\langle\cdots\rangle$ is defined as $\langle\cdots\rangle = \text{Tr}(\rho\cdots)/\text{Tr}(\rho)$, where $\rho$ denotes the density matrix at $t=0$. Associating $\langle Q(t)\rangle = x(t)$, $\langle P(t)\rangle = p(t)$, one finds that Eq. (11) is identical to Eq. (6) if $\alpha=\beta=0$. This implies that the effective Hamiltonian $H_0$ [Eq. (10)] fails to capture effects of nonzero $\alpha$ and $\beta$. In the next three sections, we attempt to resolve this problem.

### B. Caldeira-Leggett description of damping

To solve the problem, one should first find a way to describe damping. A convenient way to describe finite damping within the effective Hamiltonian approach is to adopt the Caldeira-Leggett description[32] of the damping. Its main idea is to introduce a collection of additional degrees of freedom (called environment) and couple them to the original dynamic variables so that energy of the dynamic variables can be transferred to the environment. For instance, for a one-dimensional (1D) particle subject to damped dynamics,

$$\frac{dx}{dt} = \frac{p}{M}, \qquad (12a)$$

$$\frac{dp}{dt} = -\frac{dV(x)}{dx} - \gamma\frac{dx}{dt}. \qquad (12b)$$

Caldeira and Leggett[32] demonstrated that its quantum-mechanical Hamiltonian can be constructed by adding damping Hamiltonian $H_1$ to the undamped Hamiltonian $H_0 = P^2/2M + V(Q)$. The damping Hamiltonian $H_1$ contains a collection of environmental degrees of freedom $\{x_i, p_i\}$ behaving like harmonic oscillators [see Eq. (14)], which couple to the particle through the linear coupling term $\Sigma_i C_i x_i Q$ between $Q$ and the environmental variables $x_i$. Here, $C_i$ is the coupling constant between $x_i$ and $Q$. The implication of the coupling is twofold: (i) the coupling to the environment generates damping, whose precise form depends on $C_i$, $m_i$, and $\omega_i$. It is demonstrated in Ref. 32 that the coupling generates the simple damping of the form in Eq. (12b) if $C_i$, $m_i$, and $\omega_i$ satisfy the following relation of the spectral function $J(\omega)$:

$$J(\omega) \equiv \frac{\pi}{2}\sum_i \frac{C_i^2}{m_i\omega_i}\delta(\omega - \omega_i) = \gamma\omega. \qquad (13)$$

(ii) The coupling also modified the potential $V$ by generating an additional contribution $-\Sigma_i C_i^2 Q^2/2m_i\omega_i^2$. This implies that $V(x)$ in Eq. (12b) should not be identified with $V(Q)$ in $H_0$ (even though the same symbol $V$ is used) but should be identified instead with the total potential that includes the contribution from the environmental coupling. If we express the total Hamiltonian $H$ in terms of the effective $V(x)$ that appears in Eq. (12b), it reads

$$H = H_0 + H_1, \qquad (14a)$$

$$H_0 = \frac{P^2}{2M} + V(Q), \qquad (14b)$$

$$H_1 = \sum_i \left[\frac{p_i^2}{2m_i} + \frac{1}{2}m_i\omega_i^2\left(x_i + \frac{C_i}{m_i\omega_i^2}Q\right)^2\right]. \qquad (14c)$$

By identifying $x(t) = \langle Q(t)\rangle$, $p(t) = \langle P(t)\rangle$, the equations of motion obtained from Eqs. (13) and (14) reproduce Eq. (12).

### C. Generalization to the DW motion: $\alpha$ term

Here we aim to apply the Caldeira-Leggett approach to construct an effective Hamiltonian of the DW dynamics subject to finite damping ($\alpha \neq 0$). To simplify the problem, we first focus on a situation, where only $\alpha$ is relevant and $\beta$ is irrelevant. This situation occurs if there is no current ($v_s = 0$). Then Eq. (6) reduces to

$$\frac{dx}{dt} = \frac{\alpha S}{2KM}\frac{dp}{dt} + \frac{p}{M}, \qquad (15a)$$

$$\frac{dp}{dt} = -\frac{2\alpha KM}{S}\frac{dx}{dt}. \qquad (15b)$$

Note that $\beta$ does not appear. Note also that these equations are slightly different from Eq. (12), where a damping term is contained only in the equation of $\frac{dp}{dt}$. However, in the equations of the DW [Eq. (15)], damping terms appear not only in the equation of $\frac{dp}{dt}$ [Eq. (15b)] but also in the equation of $\frac{dx}{dt}$ [Eq. (15a)].

Thus the Caldeira-Leggett description in the preceding section is not directly applicable and should be generalized. To get a hint, it is useful to recall the conjugate relation between $Q$ and $P$. The equations of $\frac{dQ}{dt}$ and $\frac{dP}{dt}$ are obtained by differentiating $H$ with respect to $P$ and $-Q$, respectively. Of course, it holds for $(x_i, p_i)$, also. Thus, one can obtain another set of Heisenberg's equation of motion by exchanging $(Q, x_i) \leftrightarrow (-P, -p_i)$. By this canonical transformation, the position coupling $\Sigma_i C_i x_i Q$ changes to a momentum coupling term, and the damping term in the equation of $\frac{dP}{dt}$ is now in that of $\frac{dQ}{dt}$. This mathematical relation that the momentum coupling generates a damping term in the equation of $\frac{dQ}{dt}$ makes it reasonable to expect that the momentum coupling $\Sigma_i D_i p_i P$ is needed[45] to generate the damping in the equation for $\frac{dQ}{dt}$. Here $D_i$ is the coupling constant between $P$ and $p_i$. The reason why, in the standard Caldeira-Leggett approach, the damping term appears only in Eq. (12b) is that Eq. (14) contains only position coupling terms $\Sigma_i C_i x_i Q$. It can be easily verified that the implications of the momentum coupling are again twofold: (i) the coupling indeed introduces the damping term in the equation of $\frac{dQ}{dt}$. (ii) it modifies the DW mass. The mass renormalization arises from the fact that in the presence of the momentum coupling $\Sigma_i D_i p_i P$, the kinematic momentum $m_i \frac{dx_i}{dt}$ of an environmental degree of freedom $x_i$ is now given by $(p_i + D_i m_i P)$ instead of $p_i$. Then the term $\Sigma_i [\frac{p_i^2}{2m_i} + D_i p_i P]$ can be decomposed into two pieces, $\Sigma_i \frac{(p_i + D_i m_i P)^2}{2m_i}$, which is the kinetic energy associated with $x_i$, and $[-\Sigma_i \frac{D_i^2 m_i}{2}]P^2$. Note that the second piece has the same form as the DW kinetic term $\frac{P^2}{2M}$. Thus this second piece generates the renormalization of the DW mass. Due to this mass renormalization effect, $M$ in Eq. (15) should be interpreted as the renormalized mass that contains the contribution from the environmental coupling. If $M$ in $H_0$ in Eq. (10) is interpreted as the renormalized mass, the environment Hamiltonian $H_2$ for the DW dynamics becomes

$$H_2 = \sum_i \left[\frac{1}{2m_i}(p_i + D_i m_i P)^2 + \frac{1}{2}m_i\omega_i^2\left(x_i + \frac{C_i}{m_i\omega_i^2}Q\right)^2\right]. \qquad (16)$$

Here, $\Sigma_i (p_i + D_i m_i P)^2/2m_i$ coupling is equivalent to the original form $\Sigma_i (p_i^2/2m_i + D_i p_i P)$ under the mass renormalization $1/M \to 1/M - \Sigma_i D_i^2 m_i/2$. Note that in $H_2$, the collective coordinates $Q$ and $P$ of the DW couple to the environmental degrees of freedom $\{x_i, p_i\}$ through two types of coupling, $\Sigma_i C_i x_i Q$ and $\Sigma_i D_i p_i P$.

Finally, one obtains the total Hamiltonian describing the DW motion in the absence of the current,

$$H = H_0|_{v_s=0} + H_2 = \frac{P^2}{2M} + \sum_i \left[ \frac{1}{2m_i}(p_i + D_i m_i P)^2 + \frac{1}{2} m_i \omega_i^2 \left( x_i + \frac{C_i}{m_i \omega_i^2} Q \right)^2 \right]. \quad (17)$$

Now, the renormalized mass $M$ in the above equation is identical to the mass in Eq. (15). To make the physical meaning of $x_i$ clearer, we perform the canonical transformation,

$$x_i \to -\frac{C_i}{m_i \omega_i^2} x_i, \quad p_i \to -\frac{m_i \omega_i^2}{C_i} p_i. \quad (18)$$

Defining $\gamma_i = \frac{C_i D_i}{\omega_i^2}$, and redefining a new $m_i$ as $m_i(\text{new}) = \frac{C_i^2}{m_i \omega_i^4}$, the Hamiltonian becomes simpler as

$$H = \frac{P^2}{2M} + \sum_i \left[ \frac{1}{2m_i}(p_i - \gamma_i P)^2 + \frac{1}{2} m_i \omega_i^2 (x_i - Q)^2 \right]. \quad (19)$$

Now, the translational symmetry of the system and the physical meaning of $x_i$ become obvious.

The next step is to impose proper constraints on $\gamma_i$ and $m_i$, so that the damping terms arising from Eq. (19) agree exactly with those in Eq. (15). For this purpose, it is convenient to introduce Laplace transformed variables $\tilde{Q}(\lambda)$, $\tilde{P}(\lambda)$, $\tilde{x}_i(\lambda)$, $\tilde{p}_i(\lambda)$, where $\tilde{Q}(\lambda) = \int_0^\infty e^{-\lambda t} \langle Q(t) \rangle dt$, and other transformed variables are defined in a similar way. Then the variables $\tilde{x}_i$ and $\tilde{p}_i$ can be integrated out easily (see Appendix B). After some tedious but straightforward algebra, it is verified that when the following three constraints on $\gamma_i$, $\omega_i$, $m_i$ are satisfied for any positive $\lambda$,

$$\sum_i \frac{\gamma_i \omega_i^2}{\lambda^2 + \omega_i^2} = 0, \quad (20a)$$

$$\sum_i \frac{\gamma_i^2 \lambda}{m_i(\lambda^2 + \omega_i^2)} = \frac{\alpha S}{2KM}, \quad (20b)$$

$$\sum_i \frac{m_i \omega_i^2 \lambda}{\lambda^2 + \omega_i^2} = \frac{2\alpha KM}{S}, \quad (20c)$$

the DW dynamics satisfies the following equation:

$$\begin{pmatrix} \lambda & -\frac{1}{M} - \frac{\alpha S \lambda}{2KM} \\ \frac{2\alpha KM}{S} \lambda & \lambda \end{pmatrix} \begin{pmatrix} \tilde{Q} \\ \tilde{P} \end{pmatrix} = \begin{pmatrix} \langle Q(0) \rangle \\ \langle P(0) \rangle \end{pmatrix} + \begin{pmatrix} -\frac{\alpha S}{2KM} \langle P(0) \rangle \\ \frac{2\alpha KM}{S} \langle Q(0) \rangle \end{pmatrix}, \quad (21)$$

which is nothing but the Laplace transformation of the DW equation [Eq. (15)] if $\langle Q \rangle$ and $\langle P \rangle$ are identified with $x$ and $p$. Thus we verify that the Hamiltonian $H$ in Eq. (19) indeed provides a generalized Caldeira-Leggett-type quantum Hamiltonian for the DW motion. As a passing remark, we mention that in the derivation of Eq. (21), the environmental degrees of freedom at the initial moment ($t=0$) are assumed to be in their thermal equilibrium so that

$$\langle x_i(0) \rangle = \langle Q(0) \rangle, \quad (22a)$$

$$\langle p_i(0) \rangle = \gamma_i \langle P(0) \rangle. \quad (22b)$$

Equation (22) can be understood as follows. First, one obtains Eq. (22) by following Appendix D which describes the statistical properties of Eq. (19) at high temperature. In Appendix D, $\langle x_i(0) - Q(0) \rangle = \langle p_i(0) - \gamma_i P(0) \rangle$ is reduced to an integration of an odd function so it is shown to vanish. The second way is probably easier to understand and does not require the classical limit or high-temperature limit. The Hamiltonian [Eq. (19)] is symmetric under the canonical transformation $Q(0) \to -Q(0)$, $P(0) \to -P(0)$, $x_i(0) \to -x_i(0)$, and $p_i(0) \to -p_i(0)$. Due to this symmetry, one obtains $\langle x_i(0) - Q(0) \rangle = \langle Q(0) - x_i(0) \rangle$ and $\langle p_i(0) - \gamma_i P(0) \rangle = \langle \gamma_i P(0) - p_i(0) \rangle$, which lead to $\langle x_i(0) \rangle = \langle Q(0) \rangle$ and $\langle p_i(0) \rangle = \gamma_i \langle P(0) \rangle$, respectively.

Here physical origin of the momentum coupling ($\gamma$) between the DW and environment deserves some discussion. Equation (19) is reduced to the original Caldeira-Leggett Hamiltonian if $\gamma_i = 0$. However, Eq. (20b) implies that the momentum coupling as well as the position coupling is indispensable to describe the Gilbert damping. To understand the origin of the momentum coupling $\gamma_i$, it is useful to recall that since $P \propto \phi \propto$ (tilting), one can interpret $P$ and $Q$ as transverse and longitudinal spin fluctuation of the DW state, respectively. (See, for explicit mathematical relation, Appendix A.) Thus, if there is rotational symmetry on spin interaction with the heat bath (or environment), the existence of the position coupling requires the existence of the momentum coupling. Thus the appearance of the damping terms both in Eqs. (15a) and (15b) is natural in view of the rotational symmetry of the spin exchange interaction and also in view of the physical meaning of $P$ and $Q$ as transverse and longitudinal spin fluctuations.

### D. Coupling with the spin current: $\beta$ term

In this section, we aim to construct a Caldeira-Leggett-type effective quantum Hamiltonian that takes account of not only $\alpha$ but also $\beta$. Since $\beta$ becomes relevant only when there exists finite spin current, we have to deal with situations with finite current ($v_s \neq 0$). Then the system is *not* in thermal equilibrium.

As demonstrated in Eq. (10), the spin current couples with the DW linear momentum, i.e., $v_s P$. Here, adiabatic velocity $v_s$ acts as the coupling constant proportional to spin current. The spin current may also couple directly to the environmental degrees of freedom. Calling this coupling constant $v$, one introduces the corresponding coupling term $\sum_i v p_i$. Later we find that this coupling is crucial to account for nonzero $\beta$. At this point we will not specify the value of $v$. Now, the total effective Hamiltonian in the presence of the spin current obtained by adding the coupling term $\sum_i v p_i$ to Eq. (19). Then,

$$H_{tot} = H + H_{current} = \frac{P^2}{2M} + v_s P + \sum_i v p_i$$
$$+ \sum_i \left[ \frac{1}{2m_i}(p_i - \gamma_i P)^2 + \frac{1}{2} m_i \omega_i^2 (x_i - Q)^2 \right]. \quad (23)$$

In order to illustrate the relation between Eqs. (6) and (23), we consider a situation, where the current is zero until $t=0$ and turned on at $t=0$ to a finite value. This situation is described by the following time-dependent Hamiltonian:

$$H_{tot} = \frac{P^2}{2M} + v_s(t) P + \sum_i v(t) p_i$$
$$+ \sum_i \left[ \frac{1}{2m_i}(p_i - \gamma_i P)^2 + \frac{1}{2} m_i \omega_i^2 (x_i - Q)^2 \right], \quad (24)$$

where $v_s(t) = v_s \Theta(t)$ and $v(t) = v \Theta(t)$. And $\Theta(t)$ is

$$\Theta(t) = \begin{cases} 1 & \text{for } t > 0, \\ 0 & \text{for } t < 0. \end{cases} \quad (25)$$

To make a quantitative comparison between Eqs. (6) and (24), one needs to integrate out environmental degrees of freedom $\{x_i, p_i\}$, which requires one to specify their initial conditions. Since the system is in thermal equilibrium until $t=0$, we may still impose the constraint in Eq. (22) to examine the DW dynamics for $t > 0$. By following a similar procedure as in Sec. II C and by using the constraints in Eq. (20),[46] one finds that the effective Hamiltonian $H$ [Eq. (24)] predicts

$$\langle Q(t) \rangle = \langle Q(0) \rangle + vt + \frac{S}{2KM\alpha^2}(1 - e^{-2K\alpha t/S(1+\alpha^2)})$$
$$\times [\alpha \langle P(0) \rangle - M\alpha(v_s - v)]. \quad (26)$$

This is exactly the same as Eq. (7) if

$$\frac{\beta}{\alpha} = \frac{v}{v_s}. \quad (27)$$

So by identifying $v$ with $v_s \beta/\alpha$, we obtain a Caldeira-Leggett-type effective quantum Hamiltonian of the DW dynamics.

One needs to consider an external force on Eq. (6b) [or Eq. (4b)] when the translational symmetry of the system is broken by some factors such as external magnetic field and magnetic defects. To describe this force, one can add a position dependent potential $V(Q)$ (Ref. 47) to Eq. (24). Considering the Heisenberg's equation, the potential $V(Q)$ generates the term $-V'(Q)$ in Eq. (6b).

### III. IMPLICATIONS

#### A. Insights on the physical meaning of $\beta$

Equation (27) provides insights on the physical meaning of $\beta$. $\beta$ depends largely on the coupling between the environment and current, not on the damping form. Recalling that $v_s$ describes the coupling between the current and the DW, we find that $\beta/\alpha$, which describes the asymptotic behavior of the DW motion, is the ratio between the current-magnetization (DW in the present case) coupling and current-environment coupling. That is,

$$\frac{\beta}{\alpha} = \frac{\text{(Coupling between the current and the environment)}}{\text{(Coupling between the current and the DW)}}. \quad (28)$$

To make the physical meaning of Eq. (28) more transparent, it is useful to examine consequences of the nonzero coupling $v$ between the current and the environment. One of the immediate consequences of the nonzero $v$ appears in the velocities of the environmental degrees of freedom. It can be verified easily that the initial velocities of environmental coordinates are given by exactly $v$, $\langle \dot{x}_i(0) \rangle = v$. Recalling that the terminal velocity of the DW, $\langle \dot{Q}(t) \rangle$ approaches $v_s \beta/\alpha$, one finds from Eq. (27) that the terminal velocity of the DW is nothing but the environment velocity. This result is very natural since the total Hamiltonian $H_{tot}$ [Eq. (23)] is Galilean invariant and the total mass of the environment (or reservoir) is much larger than the DW mass.[48] A very similar conclusion is obtained by Garate et al.[29] By analyzing the Kambersky mechanism,[49] which is reported[50] to be the dominant damping mechanism in transition metals such as Fe, Co, Ni, they found that the ratio $\beta/\alpha$ is approximately given by the ratio between the drift velocity of the Kohn-Sham quasiparticles and $v_s$. Since the collection of Kohn-Sham quasiparticles play the role of the environment in case of the Kambersky mechanism, the result in Ref. 29 is consistent with ours. It is interesting to note that our calculation, which is largely independent of details of damping mechanism, reproduces the result for the specific case.[29] This implies that the result in Ref. 29 can be generalized if the drift velocity of the Kohn-Sham quasiparticles is replaced by the general coupling constant $v$ between the current and the environment.

Our claim that the origin of $\beta$ is the direct coupling between the current and environment has an interesting conceptual consistency with the work by Zhang and Li.[4] Zhang and Li derived the nonadiabatic term by introducing a spin-relaxation term in the equation of motion of the conduction electrons. A clear consistency arises from generalizing the spin relaxation in Ref. 4 to the coupling with environment in our work. In Ref. 4, Gilbert damping ($\alpha$) and the nonadiabatic STT ($\beta$) are identified as the spin relaxation of magnetization and conduction electrons, respectively. Generalizing the spin relaxation to environmental coupling, $\alpha$ and $\beta$ are now identified as the coupling of the environment with the magnetization (i.e., the DW in our model) and the coupling of the environment with current, respectively. It is exactly how we identified $\alpha$ and $\beta$, and this gives the conceptual consistency between our work and Ref. 4. As an additional comment, while some magnitudes and origins of $\beta$ claimed in different references, such as Refs. 4 and 29, seem to be based on completely independent phenomena, our work and interpretation on $\beta$ provide a connection between them through the environmental degrees of freedom.

### B. Effect of environment on stochastic forces

Until now, our considerations has been limited to the evolution of the expectation values $\langle Q(t) \rangle$ and $\langle P(t) \rangle$ and thus thermal fluctuation effects have been ignored. In this section, we address the issue of thermal fluctuations. For this purpose, we need to go beyond the expectation values and so we derive the following *operator* equations from the Hamiltonian Eq. (23):

$$\dot{Q} = v_s + \frac{\alpha S}{2KM}\dot{P} + \frac{P}{M} + \eta_1(t), \quad (29a)$$

$$\dot{P} = -\frac{2\alpha KM}{S}\dot{Q} + \frac{2\alpha KM}{S}v + \eta_2(t), \quad (29b)$$

where

$$\eta_1(t) = \sum_i \gamma_i \omega_i \left( \Delta x_i \sin \omega_i t - \frac{\Delta p_i}{m_i \omega_i} \cos \omega_i t \right), \quad (30a)$$

$$\eta_2(t) = \sum_i (m_i \omega_i^2 \Delta x_i \cos \omega_i t + \omega_i \Delta p_i \sin \omega_i t). \quad (30b)$$

Here $\Delta x_i \equiv x_i(0) - Q(0)$ and $\Delta p_i \equiv p_i(0) - \gamma_i P(0)$. The derivation of Eqs. (29) and (30) utilizes constraints Eqs. (20) and (27). We remark that the result in Sec. II D can be recovered from Eqs. (29) and (30) by taking the expectation values of the operators. When Eq. (29) is compared to Eq. (6), it is evident that $\eta_1(t)$ and $\eta_2(t)$ defined in Eq. (30) carry the information about the thermal noise. It is easy to verify that the expectation values of $\eta_1(t)$ and $\eta_2(t)$ vanish, thus reproducing the results in the earlier section. Here it should be noticed that Eq. (30) relates $\eta_1(t)$ and $\eta_2(t)$ in the nonequilibrium situations (after the current is turned on or $t>0$) to the operators $\Delta x_i$ and $\Delta p_i$, which are defined in the equilibrium situation (right before the current is turned on or $t=0$). Thus by combining Eq. (30) with the equilibrium noise characteristics of $\Delta x_i$ and $\Delta p_i$, we can determine the thermal noise characteristic in the nonequilibrium situation ($t>0$).

To extract information about the noise, one needs to evaluate the correlation functions $\langle \{\eta_i(t), \eta_j(t)\} \rangle$ ($i,j=1,2$), where $\{,\}$ denotes the anticommutator. Due to the relations in Eq. (30), the evaluation of the correlation function reduces to the expectation value evaluation of the operator products $\{x_i(0), p_j(0)\}$, $x_i(0)x_j(0)$, and $p_i(0)p_j(0)$ in the equilibrium situation governed by the equilibrium Hamiltonian [Eq. (19)].

In the classical limit ($\hbar \to 0$, see the next paragraph to find out when the classical limit is applicable), Eq. (19) is just a collection of independent harmonic oscillators of $\{\Delta x_i, \Delta p_i\}$. Hence, the equipartition theorem determines their correlations,

$$\langle \Delta x_i \rangle = \langle \Delta p_i \rangle = \langle \Delta x_i \Delta p_i \rangle = 0, \quad (31a)$$

$$\langle \Delta x_i \Delta x_j \rangle = \frac{k_B T}{m_i \omega_i^2}\delta_{ij}, \quad (31b)$$

$$\langle \Delta p_i \Delta p_j \rangle = m_i k_B T \delta_{ij}. \quad (31c)$$

Equation (20) and (31) give the correlations of $\eta_1(t)$ and $\eta_2(t)$. After some algebra, one straightforwardly gets

$$\langle \eta_i(t) \rangle = 0, \quad (32a)$$

$$\langle \eta_1(t)\eta_2(t') \rangle = 0, \quad (32b)$$

$$\langle \eta_1(t)\eta_1(t') \rangle = \frac{\alpha S}{2KM} k_B T \delta(t-t'), \quad (32c)$$

$$\langle \eta_2(t)\eta_2(t') \rangle = \frac{2\alpha KM}{S} k_B T \delta(t-t'). \quad (32d)$$

These relations are consistent with Eq. (5) when $\eta_1(t)$ and $\eta_2(t)$ in Eq. (30) are identified with those in Eq. (6). Thus they confirm that the relations [Eq. (32)] assumed in many papers[9–13] indeed hold rather generally in the regime where the tilting angle remains sufficiently smaller than $\pi/4$.

Next we consider the regime where the condition of the classical limit is valid. Since statistical properties of the system at finite temperature is determined by $\frac{k_B T}{\hbar}$, the classical limit ($\hbar \to 0$) is equivalent to the high-temperature limit ($T \to \infty$). Thus, in actual experimental situations, the above correlation relations, Eq. (32), will be satisfied at high temperature. In this respect, we find that most experimental situations belong to the high-temperature regime. See Appendix D for the estimation of the "threshold" temperature, above which Eq. (32) is applicable. In Appendix D, the correlations in the high temperatures are derived more rigorously.

Finally we comment briefly on the low-temperature quantum regime. In this regime, one cannot use the equipartition theorem since the system is not composed of independent harmonic oscillators, that is, $[\Delta x_i, \Delta p_j] = i\hbar(\delta_{ij} + \gamma_j)$. Note that the commutator contains an additional term $i\hbar \gamma_j$. Here, the additional term $i\hbar \gamma_j$ comes from the commutator $[-Q, -\gamma_j P]$. Then, Eq. (32), which is assumed in other papers,[9–13] is not guaranteed any more.

### IV. CONCLUSION

In this paper, we examine the effect of finite current on thermal fluctuation of current-induced DW motion by constructing generalized Caldeira-Leggett-type Hamiltonian of the DW dynamics, which describes not only energy-conserving dynamics processes but also the Gilbert damping and STT. Unlike the classical damping worked out by Caldeira and Leggett,[32] the momentum coupling is indispensable to describe the Gilbert damping. This is also related to the rotational symmetry of spin-interaction nature. It is demonstrated that the derived Caldeira-Leggett-type quantum-mechanical Hamiltonian reproduces the well-known DW equations of motion.

Our Hamiltonian also illustrates that the nonadiabatic STT is closely related with the coupling of the spin current to the environment. Thus, the environmental degrees of freedom are responsible for both the Gilbert damping ($\alpha$) and the nonadiabatic STT ($\beta$). By this process, the ratio of $\beta$ and $\alpha$ was derived to be the ratio of current-DW coupling and current-environment coupling. The nonadiabatic term is nothing but the result of the direct coupling between the current and environment in our theory.

By using the Calderia-Leggett-type Hamiltonian, which describes the time evolution of the system, we obtained the expression of stochastic forces caused by thermal noise in the presence of the finite current. By calculating the equilibrium thermal fluctuation at high temperature, we verify that when $j_p$ is sufficiently smaller than the intrinsic critical density, $j_p$ does not modify the correlation relations of thermal noise unless the temperature is extremely low. The upper bound of the critical temperature, below which the above conclusion does not apply, is obtained by reexamining the system with Feynman path integral. The bound is much lower than the temperature in most experimental situations.

Lastly we remark that the Joule heating[51] is an important factor that affects the thermal fluctuation field since it raises the temperature of the nanowire. The degree of the temperature rise depends on the thermal conductivities and heat capacities of not only the nanowire but also its surrounding materials such as substrate layer materials of the nanowire. Such factors are not taken into account in this paper. Simultaneous account of the Joule heating dynamics and the thermal fluctuation field (in the presence of current) goes beyond the scope of the paper and may be a subject of future research.

### ACKNOWLEDGMENTS

We acknowledge critical comment by M. Stiles, who pointed out the importance of the momentum coupling and informed us of Ref. 29. This work was financially supported by the NRF (Grants No. 2007-0055184, No. 2009-0084542, and No. 2010-0014109) and BK21. K.W.K. acknowledges the financial support by the TJ Park.

### APPENDIX A: EFFECTIVE HAMILTONIAN OF THE DW MOTION FROM 1D s-d MODEL (Ref. 52)

The starting point is 1D $s$-$d$ model,

$$H_{s\text{-}d} = -J\sum_i \vec{S}_i \cdot \vec{S}_{i+1} - A\sum_i (\vec{S}_i \cdot \hat{z})^2 + K\sum_i (\vec{S}_i \cdot \hat{y})^2 + H_{cS}, \tag{A1}$$

as mentioned in Sec. II A.

In order to consider the DW dynamics, one first introduce the classical DW profile initially given by

$$\langle \vec{S}_i \cdot \hat{x} \rangle = S \sin \theta(z_i), \tag{A2a}$$

$$\langle \vec{S}_i \cdot \hat{y} \rangle = 0, \tag{A2b}$$

$$\langle \vec{S}_i \cdot \hat{z} \rangle = S \cos \theta(z_i), \tag{A2c}$$

where $z_i$ is the position of the $i$th localized spin, and $\theta(z) = 2\cot^{-1} e^{-\sqrt{2A/Ja^2}(z-q)}$. Here $q$ is the classical position of the DW. Small quantum fluctuations of spins on top of the classical DW profile can be described by the Holstein-Primakoff boson operator $b_i$, to describe magnon excitations. Kim et al.[43] found eigenmodes of these quantum fluctuations in the presence of the classical DW background, which amount to quantum mechanical version of the classical vibration eigenmodes in the presence of the DW background reported long time ago by Winter.[53] The corresponding eigenstates of this Hamiltonian are composed of spin-wave states with the finite eigenenergy $E_k = \sqrt{(JSa^2k^2 + 2AS)(JSa^2k^2 + 2AS + 2KS)}$ ($\geq 2S\sqrt{A(A+K)}$) and so-called bound magnon states with zero energy $E_w = 0$. Here, $k$ is the momentum of spin wave states and $a$ is the lattice spacing between two neighboring spins. Let $a_k$ and $b_w$ denote proper linear combinations of $b_i$ and $b_i^\dagger$, which represent the boson annihilation operators of finite-energy spin-wave states and zero-energy bound magnon states, respectively. In terms of these operators, Eq. (8) reduces to

$$H_{s\text{-}d} = \frac{P^2}{2M} + \sum_k E_k a_k^\dagger a_k + H_{cS}, \tag{A3}$$

where higher-order processes describing magnon-magnon interactions are ignored. Here $M$ is the so-called Döring mass,[44] defined as $M = \frac{\hbar^2}{K}\sqrt{\frac{2A}{Ja^4}}$, and $P$ is defined as $-i\hbar(\frac{2AS^2}{Ja^4})^{1/4}(b_w^\dagger - b_w)$. According to Ref. 43, $P$ is a translation generator of the DW position, that is, $\exp(iPq_0/\hbar)$ shifts the DW position by $q_0$. Thus $P$ can be interpreted as a canonical momentum of the DW translational motion. The first term in Eq. (A3), which amounts to the kinetic energy of the DW translational motion, implies that $M$ is the DW mass. We identify this $M$ with the undetermined constant $M$ in Eq. (6). According to Ref. 43, $P$ is also proportional to the degree of the DW tilting, that is, $(b_w^\dagger - b_w) \propto S_{iy}$.

In the adiabatic limit, that is, when the DW width $\lambda$ is sufficiently large in view of the electron dynamics, the remaining term $H_{cS}$ can be represented in a simple way in terms of the bound magnon operators and the adiabatic velocity of the DW,[20,43]

$$H_{cS} = v_s P. \tag{A4}$$

Then the effective $s$-$d$ Hamiltonian of the DW motion becomes

$$H_{s\text{-}d} = \frac{P^2}{2M} + v_s P + \sum_k E_k a_k^\dagger a_k. \tag{A5}$$

Note that the bound magnon part and the spin-wave part are completely decoupled in Eq. (A5) since $P$ contains only the bound magnon operators, which commute with the spin-wave operators.

The DW position operator should satisfy the following two properties: geometrical relation $\langle Q \rangle - q = \frac{a}{2S}\sum_i \langle \vec{S}_i \cdot \hat{z} \rangle$ and canonical relation $[Q,P]=i\hbar$. Then, one can show that $Q=q-(\frac{Ja^4}{32AS^2})^{1/4}(b_w^\dagger+b_w)$ satisfies these two properties. Note that $Q$ is expressed in terms of the bound magnon operators. Then as far as the Heisenberg equations of motion for $Q$ and $P$ are concerned, the last term in Eq. (A5) does not play any role. This term will be ignored from now on. Thus, the effective Hamiltonian for the DW motion is reduced to

$$H_0 = \frac{P^2}{2M} + v_s P \quad \text{(A6)}$$

so we got Eq. (10), the effective Hamiltonian of the DW motion.

## APPENDIX B: SOLUTION FOR A GENERAL QUADRATIC DAMPING

This section provides the solution of the equation of motion for a general quadratic damping. This is applicable not only for the generalized Caldeira-Leggett description in this paper but also for any damping type which quadratically interacts with the DW.

In general, let us consider a general quadratic damping Hamiltonian,

$$H = \frac{P^2}{2M} + v_s P + \sum_i \chi_i^T A^i \chi_i, \quad \text{(B1)}$$

where $\chi_i=(Q\ P\ x_i\ p_i)^T$, and $A^i$ is a $4\times 4$ Hermitian matrix. Now, one straightforwardly gets the corresponding coupled equations,

$$\frac{dQ}{dt} = \frac{P}{M} + v_s + \sum_i (B_{21}^i Q + B_{22}^i P + B_{23}^i x_i + B_{24}^i p_i), \quad \text{(B2a)}$$

$$\frac{dP}{dt} = -\sum_i (B_{11}^i Q + B_{12}^i P + B_{13}^i x_i + B_{14}^i p_i), \quad \text{(B2b)}$$

$$\frac{dx_i}{dt} = B_{41}^i Q + B_{42}^i P + B_{43}^i x_i + B_{44}^i p_i, \quad \text{(B2c)}$$

$$\frac{dp_i}{dt} = -(B_{31}^i Q + B_{32}^i P + B_{33}^i x_i + B_{34}^i p_i). \quad \text{(B2d)}$$

Here, $B^i$ is a $4\times 4$ real symmetric matrix defined as $B^i = 2\operatorname{Re}[A^i]$, and $B_{jk}^i$ is the element of $B^i$ in $j$th row and $k$th column.

With the Laplace transform of the expectation values of each operator, for example,

$$\tilde{Q}(\lambda) \equiv \mathcal{L}[Q(t)](\lambda) = \int_0^\infty \langle Q(t)\rangle e^{-\lambda t} dt, \quad \text{(B3)}$$

the set of coupled equations transforms as

$$\lambda \tilde{Q} - \langle Q(0)\rangle = \frac{\tilde{P}}{M} + \frac{v_s}{\lambda} + \sum_i (B_{21}^i \tilde{Q} + B_{22}^i \tilde{P} + B_{23}^i \tilde{x}_i + B_{24}^i \tilde{p}_i), \quad \text{(B4a)}$$

$$\lambda \tilde{P} - \langle P(0)\rangle = -\sum_i (B_{11}^i \tilde{Q} + B_{12}^i \tilde{P} + B_{13}^i \tilde{x}_i + B_{14}^i \tilde{p}_i), \quad \text{(B4b)}$$

$$\lambda \tilde{x}_i - \langle x_i(0)\rangle = B_{41}^i \tilde{Q} + B_{42}^i \tilde{P} + B_{43}^i \tilde{x}_i + B_{44}^i \tilde{p}_i, \quad \text{(B4c)}$$

$$\lambda \tilde{p}_i - \langle p_i(0)\rangle = -(B_{31}^i \tilde{Q} + B_{32}^i \tilde{P} + B_{33}^i \tilde{x}_i + B_{34}^i \tilde{p}_i). \quad \text{(B4d)}$$

Rewriting these in matrix forms, the equations become simpler as

$$\lambda \begin{pmatrix} \tilde{Q} \\ \tilde{P} \end{pmatrix} - \begin{pmatrix} \langle Q(0)\rangle + \frac{v_s}{\lambda} \\ \langle P(0)\rangle \end{pmatrix} = \left\{ \begin{pmatrix} 0 & \frac{1}{M} \\ 0 & 0 \end{pmatrix} + \sum_i \begin{pmatrix} B_{21}^i & B_{22}^i \\ -B_{11}^i & -B_{12}^i \end{pmatrix} \right\} \\ \times \begin{pmatrix} \tilde{Q} \\ \tilde{P} \end{pmatrix} + \sum_i \begin{pmatrix} B_{23}^i & B_{24}^i \\ -B_{13}^i & -B_{14}^i \end{pmatrix} \\ \times \begin{pmatrix} \tilde{x}_i \\ \tilde{p}_i \end{pmatrix}, \quad \text{(B5a)}$$

$$\lambda \begin{pmatrix} \tilde{x}_i \\ \tilde{p}_i \end{pmatrix} - \begin{pmatrix} \langle x_i(0)\rangle \\ \langle p_i(0)\rangle \end{pmatrix} = \begin{pmatrix} B_{41}^i & B_{42}^i \\ -B_{31}^i & -B_{32}^i \end{pmatrix} \begin{pmatrix} \tilde{Q} \\ \tilde{P} \end{pmatrix} \\ + \begin{pmatrix} B_{43}^i & B_{44}^i \\ -B_{33}^i & -B_{34}^i \end{pmatrix} \begin{pmatrix} \tilde{x}_i \\ \tilde{p}_i \end{pmatrix}. \quad \text{(B5b)}$$

From Eq. (B5b), one can calculate $(\tilde{x}_i\ \tilde{p}_i)^T$ in terms of $\tilde{Q}$ and $\tilde{P}$,

$$\begin{pmatrix} \tilde{x}_i \\ \tilde{p}_i \end{pmatrix} = \begin{pmatrix} \lambda - B_{43}^i & -B_{44}^i \\ B_{33}^i & \lambda + B_{34}^i \end{pmatrix}^{-1} \begin{pmatrix} B_{41}^i & B_{42}^i \\ -B_{31}^i & -B_{32}^i \end{pmatrix} \begin{pmatrix} \tilde{Q} \\ \tilde{P} \end{pmatrix} \\ + \begin{pmatrix} \lambda - B_{43}^i & -B_{44}^i \\ B_{33}^i & \lambda + B_{34}^i \end{pmatrix}^{-1} \begin{pmatrix} \langle x_i(0)\rangle \\ \langle p_i(0)\rangle \end{pmatrix}. \quad \text{(B6)}$$

From Eqs. (B5a) and (B6), one finally gets the equation of $(\tilde{Q}\tilde{P})^T$,

$$\left[ \left( \begin{matrix} \lambda & -\frac{1}{M} \\ 0 & \lambda \end{matrix} \right) - \sum_i \left\{ \left( \begin{matrix} B_{21}^i & B_{22}^i \\ -B_{11}^i & -B_{12}^i \end{matrix} \right) + \left( \begin{matrix} B_{23}^i & B_{24}^i \\ -B_{13}^i & -B_{14}^i \end{matrix} \right) \left( \begin{matrix} \lambda - B_{43}^i & -B_{44}^i \\ B_{33}^i & \lambda + B_{34}^i \end{matrix} \right)^{-1} \left( \begin{matrix} B_{41}^i & B_{42}^i \\ -B_{31}^i & -B_{32}^i \end{matrix} \right) \right\} \right] \left( \begin{matrix} \widetilde{Q} \\ \widetilde{P} \end{matrix} \right)$$
$$= \left( \begin{matrix} \langle Q(0) \rangle + \frac{v_s}{\lambda} \\ \langle P(0) \rangle \end{matrix} \right) + \sum_i \left( \begin{matrix} B_{23}^i & B_{24}^i \\ -B_{13}^i & -B_{14}^i \end{matrix} \right) \left( \begin{matrix} \lambda - B_{43}^i & -B_{44}^i \\ B_{33}^i & \lambda + B_{34}^i \end{matrix} \right)^{-1} \left( \begin{matrix} \langle x_i(0) \rangle \\ \langle p_i(0) \rangle \end{matrix} \right). \quad \text{(B7)}$$

Inverting the matrix in front of $(\widetilde{Q}\widetilde{P})^T$, one can get the solution of $(\widetilde{Q}\widetilde{P})^T$. Then, finally, the solution $(\langle Q \rangle \langle P \rangle)^T$ is obtained by the inverse Laplace transform of $(\widetilde{Q}\widetilde{P})^T$,

$$\left( \begin{matrix} \langle Q(t) \rangle \\ \langle P(t) \rangle \end{matrix} \right) = \mathcal{L}^{-1} \left[ \left( \begin{matrix} \widetilde{Q}(\lambda) \\ \widetilde{P}(\lambda) \end{matrix} \right) \right]. \quad \text{(B8)}$$

### APPENDIX C: SOLUTION OF EQ. (24)

In the special case that current is applied at $t=0$, $\Theta(t)$ in Eq. (25) becomes Heaviside step function. This is the case we are interested in. In a real DW system, the DW velocity jumps from 0 to a finite value at the moment that the spin current starts to be applied. This jumping comes from the discontinuity in Eq. (25) which makes the Hamiltonian discontinuous. Right before the current is applied, the DW remains on the stable (or equilibrium) state described by Eq. (22).

Suppose that Eq. (20a) also holds for $\lambda=0$. Then, Eq. (24) transforms as (up to constant)

$$H_{tot} = \frac{P^2}{2M} + v_s(t)P + \sum_i \left[ \frac{1}{2m_i} (p_i - \gamma_i P + m_i v(t))^2 \right]$$
$$+ \sum_i \frac{1}{2} m_i \omega_i^2 (x_i - Q)^2. \quad \text{(C1)}$$

Performing the canonical transform $p_i \to p_i - m_i v(t)$, one can transform this Hamiltonian in the form of Eq. (B1),

$$H_{tot} = \frac{P^2}{2M} + v_s(t)P + \sum_i \left[ \frac{1}{2m_i} (p_i - \gamma_i P)^2 + \frac{1}{2} m_i \omega_i^2 (x_i - Q)^2 \right].$$

Here, one of the constraints Eq. (20a) is generalized to hold even for $\lambda=0$, so that $\Sigma_i \gamma_i = 0$. Note that the discontinuity due to $v(t)$ is absorbed in the new $p_i$. Thus, Eq. (22b) should be written as

$$\langle p_i(0+) \rangle = \langle p_i(0-) \rangle + m_i v = \gamma_i \langle P(0) \rangle + m_i v. \quad \text{(C2)}$$

The initial condition of $x_i$ is the same as Eq. (22a). Now, using these initial conditions and Eqs. (B7) and (B8) under the constraints in Eq. (20), one gets the solution of this system as Eq. (26).

### APPENDIX D: CORRELATIONS OF STOCHASTIC FORCES AT HIGH TEMPERATURE

This section provides the quantum derivation of correlation relations of stochastic forces at high temperature. The classical correlation relations in Eq. (32) are valid quantum mechanically at high temperature. Since Eq. (31) implies Eq. (32), it suffices to show Eq. (31) in this section. The basic strategy is studying statistical properties of the Hamiltonian Eq. (19) (under quadratic potential $bQ^2$)[54] by the Feynman path integral along the imaginary-time axis. The Feynman path integral of a system described by a quadratic Lagrangian is proportional to the exponential of the action value evaluated at the classical solution. Hence, the key point of the procedure is to get the classical solution with imaginary time.

#### 1. General relations

##### a. Classical action under high-temperature limit

Define a column vector $\chi = (Qx_1x_2\cdots)^T$. Let the Euclidean Lagrangian of the system be $L_E = \frac{1}{2}\dot{\chi}^T A \dot{\chi} + \frac{1}{2}\chi^T B\chi$, where $A$ and $B$ are symmetric matrices. (The symbols "$A$" and "$B$" are not the same as those in Appendix B.) Explicitly, $L = \frac{1}{2}\Sigma_{nm}\dot{x}_n A_{nm}\dot{x}_m + \frac{1}{2}\Sigma_{nm}x_n B_{nm}x_m$. Here $x_0 \equiv Q$. $\frac{\partial L_E}{\partial \dot{x}_n} = \Sigma_m A_{nm}\dot{x}_m = A\dot{\chi}$ and $\frac{\partial L_E}{\partial x_n} = \Sigma_m B_{nm}x_m = B\chi$ lead to the classical equation of motion,

$$A\ddot{\chi} = B\chi. \quad \text{(D1)}$$

The classical action value $S_c$ (evaluated at the classical path) is then, $S_c = \int_0^\tau L_E dt = \frac{1}{2}\int_0^\tau (\dot{\chi}^T A\dot{\chi} + \chi^T B\chi)dt = \frac{1}{2}\chi^T A\dot{\chi}|_0^\tau + \int_0^\tau (-\chi^T A\ddot{\chi} + \chi^T B\chi)dt = \frac{1}{2}\chi^T A\dot{\chi}|_0^\tau$. Here, $\tau = \hbar/k_B T$. Now, the only thing one needs is to find $\dot{\chi}$ at boundary points.

In the case of Eq. (19), $A$ is invertible. Hence, the equation becomes $\ddot{\chi} = A^{-1}B\chi$. Suppose that $A^{-1}B$ is diagonalizable, that is $A^{-1}B = C^{-1}DC$. Here $D_{nm} = \lambda_n \delta_{nm}$ is diagonal matrix and $\lambda_n$ is $n$th eigenvalue of $A^{-1}B$. Define a new vector $\xi = C\chi$. Finally, we get the equation,

$$\ddot{\xi} = \left( \begin{matrix} \lambda_0 & 0 & \cdots \\ 0 & \lambda_1 & \cdots \\ \vdots & \vdots & \ddots \end{matrix} \right) \xi. \quad \text{(D2)}$$

Imposing the boundary condition $\chi(0) = \chi_i$, $\chi(\tau) = \chi_f$ and defining the corresponding $\xi_i = C\chi_i$, $\xi_f = C\chi_f$, then one gets the solution of $\xi$ and its derivative straightforwardly,

$$\xi_n = \frac{\xi_{fn} + \xi_{in}}{2} \frac{\cosh \sqrt{\lambda_n}\left(t - \frac{\tau}{2}\right)}{\cosh \sqrt{\lambda_n}\frac{\tau}{2}} + \frac{\xi_{fn} - \xi_{in}}{2} \frac{\sinh \sqrt{\lambda_n}\left(t - \frac{\tau}{2}\right)}{\sinh \sqrt{\lambda_n}\frac{\tau}{2}}, \quad (D3)$$

$$\dot{\xi}_n = \sqrt{\lambda_n} \left[ \frac{\xi_{fn} + \xi_{in}}{2} \frac{\sinh \sqrt{\lambda_n}\left(t - \frac{\tau}{2}\right)}{\cosh \sqrt{\lambda_n}\frac{\tau}{2}} + \frac{\xi_{fn} - \xi_{in}}{2} \frac{\cosh \sqrt{\lambda_n}\left(t - \frac{\tau}{2}\right)}{\sinh \sqrt{\lambda_n}\frac{\tau}{2}} \right]. \quad (D4)$$

Now, $\dot{\xi}$ at boundary points are obtained as

$$\dot{\xi}_n(0) = \sqrt{\lambda_n}\left(-\frac{\xi_{fn} + \xi_{in}}{2}\tanh\sqrt{\lambda_n}\frac{\tau}{2} + \frac{\xi_{fn} - \xi_{in}}{2}\coth\sqrt{\lambda_n}\frac{\tau}{2}\right), \quad (D5)$$

$$\dot{\xi}_n(\tau) = \sqrt{\lambda_n}\left(\frac{\xi_{fn} + \xi_{in}}{2}\tanh\sqrt{\lambda_n}\frac{\tau}{2} + \frac{\xi_{fn} - \xi_{in}}{2}\coth\sqrt{\lambda_n}\frac{\tau}{2}\right). \quad (D6)$$

If $\frac{\sqrt{|\lambda_n|}\tau}{2} = \frac{\sqrt{|\lambda_n|}\hbar}{2k_BT} \ll 1$, $\tanh\frac{\sqrt{\lambda_n}\tau}{2} \approx \frac{\sqrt{\lambda_n}\tau}{2}$. Then,

$$\dot{\xi}_n(0) \approx -\frac{\xi_{fn} + \xi_{in}}{2}\frac{\lambda_n\tau}{2} + \frac{\xi_{fn} - \xi_{in}}{\tau}, \quad (D7)$$

$$\dot{\xi}_n(\tau) \approx \frac{\xi_{fn} + \xi_{in}}{2}\frac{\lambda_n\tau}{2} + \frac{\xi_{fn} - \xi_{in}}{\tau}. \quad (D8)$$

In matrix form,

$$\dot{\xi}(0) \approx -D\frac{\xi_f + \xi_i}{2}\frac{\tau}{2} + \frac{\xi_f - \xi_i}{\tau} = -DC\frac{\chi_f + \chi_i}{2}\frac{\tau}{2} + C\frac{\chi_f - \chi_i}{\tau}, \quad (D9)$$

$$\dot{\xi}(\tau) \approx D\frac{\xi_f + \xi_i}{2}\frac{\tau}{2} + \frac{\xi_f - \xi_i}{\tau} = DC\frac{\chi_f + \chi_i}{2}\frac{\tau}{2} + C\frac{\chi_f - \chi_i}{\tau}. \quad (D10)$$

Using $A^{-1}B = C^{-1}DC$, it leads to

$$\dot{\chi}(0) \approx -A^{-1}B\frac{\chi_f + \chi_i}{2}\frac{\tau}{2} + \frac{\chi_f - \chi_i}{\tau}, \quad (D11)$$

$$\dot{\chi}(\tau) \approx A^{-1}B\frac{\chi_f + \chi_i}{2}\frac{\tau}{2} + \frac{\chi_f - \chi_i}{\tau}. \quad (D12)$$

Finally one can obtain the classical action,

$$S_c = \frac{1}{2}\chi^T A\dot{\chi}|_0^\tau = \left(\frac{\chi_f + \chi_i}{2}\right)^T B\left(\frac{\chi_f + \chi_i}{2}\right)\frac{\tau}{2}$$
$$+ \left(\frac{\chi_f - \chi_i}{2}\right)^T A\left(\frac{\chi_f - \chi_i}{2}\right)\frac{2}{\tau}. \quad (D13)$$

This is valid even if some eigenvalues are zero. (By taking limit of $\lambda_i \to 0$, cosh and sinh becomes constant and linear, respectively.)

### b. Propagator and its derivatives

The propagator is given by the Feynman path integral, $K(\chi_f, \chi_i; \tau) = \langle \chi_f | e^{-H/k_BT} | \chi_i \rangle = \int \mathcal{D}\chi e^{-\int L_E dt/\hbar}$, where $\mathcal{D}\chi = \Pi_i \mathcal{D}x_i$. For quadratic Lagrangian, it is well known that $\int \mathcal{D}\chi e^{-\int L_E dt/\hbar} = F(\tau)e^{-S_c/\hbar}$. Here $F(\tau)$ is a smooth function dependent on $\tau$ only.

Now we aim to calculate $K(\chi_i + \delta\chi, \chi_i; \tau)$. It is easy to obtain the corresponding classical action by replacing $\chi_f = \chi_i + \delta\chi$ in Eq. (D13),

$$S_c(\chi_i + \delta\chi, \chi_i; \tau) = \frac{\tau}{2}\chi_i^T B\chi_i + \frac{\tau}{2}\chi_i^T B\delta\chi + \frac{\tau}{8}\delta\chi^T B\delta\chi$$
$$+ \frac{1}{2\tau}\delta\chi^T A\delta\chi. \quad (D14)$$

Then, $K(\chi_i + \delta\chi, \chi_i; \tau)$ is (up to second order of $\delta\chi$),

$$K(\chi_i + \delta\chi, \chi_i; \tau) = F(\tau)e^{-S_c/\hbar} = F(\tau)e^{-(\tau/2\hbar)\chi_i^T B\chi_i}$$
$$\times \left[ 1 - \frac{1}{\hbar}\left(\frac{\tau}{2}\chi_i^T B\delta\chi + \frac{\tau}{8}\delta\chi^T B\delta\chi\right.\right.$$
$$\left.\left. + \frac{1}{2\tau}\delta\chi^T A\delta\chi\right) + \frac{1}{2\hbar^2}\left(\frac{\tau}{2}\chi_i^T B\delta\chi\right)^2 \right].$$

Zeroth order: $F(\tau)e^{-(\tau/2\hbar)\chi_i^T B\chi_i}$.

First order: $-\frac{\tau}{2\hbar}F(\tau)e^{-(\tau/2\hbar)\chi_i^T B\chi_i}\chi_i^T B\delta\chi$

$$= -\frac{\tau}{2\hbar}F(\tau)e^{-(\tau/2\hbar)\chi_i^T B\chi_i}\sum_{nm} x_{in}B_{nm}\delta x_m.$$

Second order: $F(\tau)e^{-(\tau/2\hbar)\chi_i^T B\chi_i}\left\{-\frac{1}{\hbar}\left(\frac{\tau}{8}\delta\chi^T B\delta\chi\right.\right.$
$$\left.\left. + \frac{1}{2\tau}\delta\chi^T A\delta\chi\right) + \frac{1}{2\hbar^2}\left(\frac{\tau}{2}\chi_i^T B\delta\chi\right)^2 \right\}$$
$$= F(\tau)e^{-(\tau/2\hbar)\chi_i^T B\chi_i}\left\{-\frac{1}{2\hbar}\sum_{nm}\delta x_n\left(\frac{\tau}{4}B_{nm} + \frac{1}{\tau}A_{nm}\right)\delta x_m\right.$$
$$\left. + \frac{\tau^2}{8\hbar^2}\left(\sum_{klmn} x_{ik}B_{kn}\delta x_n x_{il}B_{lm}\delta x_m\right)\right\}. \quad (D15)$$

By the relation, $K(\chi_i + \delta\chi, \chi_i; \tau) = K(\chi_i, \chi_i; \tau) + \sum_m \frac{\partial K}{\partial x_{fm}}\delta x_m + \sum_{nm}\frac{1}{2}\frac{\partial^2 K}{\partial x_{fn}\partial x_{fm}}\delta x_n\delta x_m + \mathcal{O}(\delta\chi^3)$,

$$K(\chi_i, \chi_i; \tau) = F(\tau)e^{-(\tau/2\hbar)\chi_i^T B\chi_i}, \quad (D16)$$

$$\left.\frac{\partial K}{\partial x_{fm}}\right|_{\chi_i=\chi_f} = -\frac{\tau}{2\hbar}F(\tau)e^{-(\tau/2\hbar)\chi_i^T B \chi_i}\sum_n B_{nm}x_{in}, \quad (D17)$$

$$\left.\frac{\partial^2 K}{\partial x_{fn}\partial x_{fm}}\right|_{\chi_i=\chi_f} = F(\tau)e^{-(\tau/2\hbar)\chi_i^T B \chi_i}\left\{-\frac{1}{\hbar}\left(\frac{\tau}{4}B_{nm}+\frac{1}{\tau}A_{nm}\right)\right.$$
$$+\frac{\tau^2}{4\hbar^2}\left(\sum_{kl}B_{kn}x_{ik}B_{lm}x_{il}\right)\right\}$$
$$= F(\tau)e^{-(\tau/2\hbar)\chi_i^T B \chi_i}\left\{-\frac{1}{\hbar}\left(\frac{\tau}{4}B_{nm}+\frac{1}{\tau}A_{nm}\right)\right.$$
$$\left.+\frac{\tau^2}{4\hbar^2}\left(\sum_k B_{kn}x_{ik}\right)\left(\sum_k B_{km}x_{ik}\right)\right\}. \quad (D18)$$

#### c. Correlations

Statistical average of an operator $A$ is given by $\frac{\mathrm{Tr}(Ae^{-H/k_B T})}{\mathrm{Tr}(e^{-H/k_B T})}$. What we want to find are the averages of $\Delta x_n \Delta x_m$, $\Delta p_n \Delta p_m$, and $\{\Delta x_n, \Delta p_m\}$ for $\Delta x_n \equiv x_n - Q$ and $\Delta p_n \equiv p_n - \gamma_n P$,

$$\mathrm{Tr}(\Delta x_n \Delta x_m e^{-H/k_B T}) = \int d\chi_i \langle\chi_i|\Delta x_n \Delta x_m e^{-H/k_B T}|\chi_i\rangle$$
$$= \int d\chi_i(x_{in}-Q_i)(x_{im}-Q_i)\langle\chi_i|e^{-H/k_B T}|\chi_i\rangle$$
$$= \int d\chi_i(x_{in}-Q_i)(x_{im}-Q_i)K(\chi_i,\chi_i;\tau), \quad (D19)$$

$$\mathrm{Tr}(\Delta p_n \Delta p_m e^{-H/k_B T}) = \int d\chi_i \langle\chi_i|\Delta p_n \Delta p_m e^{-H/k_B T}|\chi_i\rangle$$
$$= -\hbar^2 \int d\chi_i \left(\frac{\partial}{\partial x_{fn}}-\gamma_n\frac{\partial}{\partial Q_f}\right)$$
$$\times\left.\left(\frac{\partial}{\partial x_{fm}}-\gamma_m\frac{\partial}{\partial Q_f}\right)K(\chi_f,\chi_i;\tau)\right|_{\chi_i=\chi_f}, \quad (D20)$$

$$\mathrm{Tr}(\Delta x_n \Delta p_m e^{-H/k_B T}) = \int d\chi_i \langle\chi_i|\Delta x_n \Delta p_m e^{-H/k_B T}|\chi_i\rangle$$
$$= -i\hbar \int d\chi_i(x_{in}-Q_i)\left(\frac{\partial}{\partial x_{fm}}\right.$$
$$\left.-\gamma_m\frac{\partial}{\partial Q_f}\right)K(\chi_f,\chi_i;\tau)\bigg|_{\chi_i=\chi_f}, \quad (D21)$$

$$\mathrm{Tr}(e^{-H/k_B T}) = \int d\chi_i \langle\chi_i|e^{-H/k_B T}|\chi_i\rangle = \int d\chi_i K(\chi_i,\chi_i;\tau), \quad (D22)$$

where $d\chi_i = \Pi_n dx_{in}$.

#### 2. Correlations under quadratic potential

Under potential $bQ^2$, the matrices $A$ and $B$ corresponding the Hamiltonian Eq. (19) are

$$A = \begin{pmatrix} M & M\gamma_1 & M\gamma_2 & \cdots \\ M\gamma_1 & M\gamma_1^2+m_1 & M\gamma_1\gamma_2 & \cdots \\ M\gamma_2 & M\gamma_2\gamma_1 & M\gamma_2^2+m_2 & \cdots \\ \vdots & \vdots & \vdots & \ddots \end{pmatrix}, \quad (D23)$$

$$B = \begin{pmatrix} b+\sum_n m_n\omega_n^2 & -m_1\omega_1^2 & -m_2\omega_2^2 & \cdots \\ -m_1\omega_1^2 & m_1\omega_1^2 & 0 & \cdots \\ -m_2\omega_2^2 & 0 & m_2\omega_2^2 & \cdots \\ \vdots & \vdots & \vdots & \ddots \end{pmatrix}. \quad (D24)$$

Then, $e^{-(\tau/2\hbar)\chi_i^T B \chi_i}$ is written as $e^{-(\tau/2\hbar)[\sum_n m_n\omega_n^2(Q_i-x_{in})^2+bQ_i^2]}$.

#### a. x-x correlations

Since $K(\chi,\chi;\tau)$ is an even function of $(x_n-Q_i)$, it is trivial that $\mathrm{Tr}(\Delta x_n \Delta x_m e^{-H/k_B T})=0$ unless $n=m$.
For $n=m$, $\mathrm{Tr}(\Delta x_n^2 e^{-H/k_B T})=\int d\chi_i(x_{in}-Q_i)^2 K(\chi_i,\chi_i;\tau)$. Thus,

$$\frac{\mathrm{Tr}(\Delta x_n^2 e^{-H/k_B T})}{\mathrm{Tr}(e^{-H/k_B T})} = \frac{\int dx_{in}(x_{in}-Q_i)^2 e^{-(\tau/2\hbar)m_n w_n^2(Q_i-x_{in})^2}}{\int dx_{in}e^{-(\tau/2\hbar)m_n w_n^2(Q_i-x_{in})^2}}$$
$$= \frac{\hbar}{\tau m_n w_n^2} = \frac{k_B T}{m_n w_n^2}. \quad (D25)$$

So, finally one gets $\langle \Delta x_n \Delta x_m \rangle = \frac{k_B T}{m_n w_n^2}\delta_{nm}$.

#### b. x-p correlations

Explicitly rewriting the derivative of $K$,

$$\left.\frac{\partial K}{\partial x_{fm}}\right|_{\chi_i=\chi_f} = -\frac{\tau}{2\hbar}K(\chi_i,\chi_i;\tau)m_m\omega_m^2(x_{im}-Q_i) \quad \text{for} \quad (m \neq 0), \quad (D26)$$

$$\left.\frac{\partial K}{\partial Q_f}\right|_{\chi_i=\chi_f} = -\frac{\tau}{2\hbar}K(\chi_i,\chi_i;\tau)\left\{bQ_i^2+\sum_n m_n\omega_n^2(Q_i-x_{in})\right\}. \quad (D27)$$

Using the above relations,

$$\mathrm{Tr}(\Delta x_n \Delta p_m e^{-H/k_B T}) = -i\hbar \int d\chi_i (x_{in} - Q_i)\left(\frac{\partial}{\partial x_{fm}} - \gamma_m \frac{\partial}{\partial Q_f}\right) K(\chi_f, \chi_i; \tau)\bigg|_{\chi_i = \chi_f}$$

$$= -\frac{i\tau}{2}\int d\chi_i (x_{in} - Q_i)\left\{\gamma_m b Q_i^2 + \gamma_m \sum_l m_l \omega_l^2 (Q_i - x_{il}) + m_m \omega_m^2 (Q_i - x_{im})\right\} K(\chi_i, \chi_i; \tau)$$

$$= -\frac{i\tau}{2}\int d\chi_i (x_{in} - Q_i)\left\{\gamma_m \sum_l m_l \omega_l^2 (Q_i - x_{il}) + m_m \omega_m^2 (Q_i - x_{im})\right\} K(\chi_i, \chi_i; \tau)$$

$$= \frac{i\tau}{2}\gamma_m \sum_l m_l \omega_l^2 \mathrm{Tr}(\Delta x_{in} \Delta x_{il} e^{-H/k_B T})) + m_m \omega_m^2 \mathrm{Tr}(\Delta x_{in} \Delta x_{im} e^{-H/k_B T}). \tag{D28}$$

In the third line, it is used that $\int dx_{in}(x_{in} - Q_i)\times [\text{even function of}(x_{in} - Q_i)] = 0$.

One can now write the $x$-$p$ correlations in terms of $x$-$x$ correlations.

$$\langle \Delta x_n \Delta p_m \rangle = \frac{i\tau}{2}\left(\gamma_m \sum_l m_l \omega_l^2 \langle \Delta x_{in} \Delta x_{il}\rangle + m_m \omega_m^2 \langle \Delta x_{in} \Delta x_{im}\rangle\right)$$

$$= \frac{i\tau k_B T}{2}\left(\gamma_m \sum_l \delta_{nl} + \delta_{nm}\right) = \frac{i\hbar}{2}(\gamma_m + \delta_{nm}), \tag{D29}$$

which is purely imaginary. Thus, $\langle\{\Delta x_n, \Delta p_m\}\rangle = \langle \Delta x_n \Delta p_m\rangle + \langle \Delta x_n \Delta p_m\rangle^* = 0$.

### c. p-p correlations

It is convenient to calculate $\int d\chi_i \frac{\partial^2 K}{\partial x_{fn} \partial x_{fm}}\big|_{\chi_i = \chi_f}$. The trickiest part is $\int d\chi_i \sum_k B_{kn} x_{ik} \sum_k B_{km} x_{ik} K(\chi_i, \chi_i; \tau)$,

$n \ne 0, \ m \ne 0$: $\sum_k B_{kn} x_{ik} \sum_k B_{km} x_{ik} = m_n \omega_n^2 (x_{in} - Q_i) m_m \omega_m^2 (x_{im} - Q_i),$

$n = 0, \ m \ne 0$: $\sum_k B_{kn} x_{ik} \sum_k B_{km} x_{ik} = \left(\sum_k m_k \omega_k^2 (Q_i - x_{ik}) + bQ_i\right) m_m \omega_m^2 (x_{im} - Q_i),$

$n = 0, \ m = 0$: $\sum_k B_{kn} x_{ik} \sum_k B_{km} x_{ik} = \left(\sum_k m_k \omega_k^2 (Q_i - x_{ik}) + bQ_i\right)\times\left(\sum_k m_k \omega_k^2 (Q_i - x_{ik}) + bQ_i\right).$

After integrating over $x_{ik}$, odd terms with respect to $(x_{ik} - Q)$ vanish. Taking only even terms, one obtains

$n \ne 0, \ m \ne 0 \to m_n^2 \omega_n^4 (x_{in} - Q_i)^2 \delta_{nm} = m_m \omega_m^2 (x_{in} - Q_i)^2 B_{nm},$

$n = 0, \ m \ne 0 \to -m_m^2 \omega_m^4 (Q_i - x_{im})^2 = m_m \omega_m^2 (x_{im} - Q_i)^2 B_{nm},$

$n = 0, \ m = 0 \to \sum_k m_k^2 \omega_k^4 (Q_i - x_{ik})^2 + b^2 Q_i^2.$

Integrating out and using the identity $\int du u^2 e^{-u^2/2\alpha} = \alpha \int du e^{-u^2/2\alpha}$ for $\alpha > 0$, one finds

$n \ne 0, \ m \ne 0$: $\int d\chi_i m_m \omega_m^2 (x_{in} - Q_i)^2 B_{nm} K(\chi_i, \chi_i; \tau)$

$$= \frac{\hbar}{\tau} B_{nm}\int d\chi_i K(\chi_i, \chi_i; \tau),$$

$n = 0, \ m \ne 0$: $\int d\chi_i m_m \omega_m^2 (x_{im} - Q_i)^2 B_{nm} K(\chi_i, \chi_i; \tau)$

$$= \frac{\hbar}{\tau} B_{nm}\int d\chi_i K(\chi_i, \chi_i; \tau),$$

$n = 0, \ m = 0$: $\int d\chi_i \left[\sum_k m_k^2 \omega_k^4 (Q_i - x_{ik})^2 + b^2 Q_i^2\right] K(\chi_i, \chi_i; \tau)$

$$= \frac{\hbar}{\tau}\left(\sum_k m_k \omega_k^2 + b\right)\int d\chi_i K(\chi_i, \chi_i; \tau)$$

$$= \frac{\hbar}{\tau} B_{nm}\int d\chi_i K(\chi_i, \chi_i; \tau).$$

The result is $\frac{\hbar}{\tau} B_{nm}\int d\chi_i K$ independent of the cases. Finally, one can obtain

$$\int d\chi_i \frac{\partial^2 K}{\partial x_{fn} \partial x_{fm}}\bigg|_{\chi_i = \chi_f} = \left\{-\frac{1}{\hbar}\left(\frac{\tau}{4}B_{nm} + \frac{1}{\tau}A_{nm}\right)\right.$$

$$\left. + \frac{\tau}{4\hbar}B_{nm}\right\}\int d\chi_i K(\chi_i, \chi_i; \tau)$$

$$= -\frac{A_{nm}}{\tau\hbar}\int d\chi_i K(\chi_i, \chi_i; \tau), \tag{D30}$$

or equivalently,

$$\left\langle \frac{\partial^2}{\partial x_n \partial x_m} \right\rangle = -\frac{k_B T}{\hbar^2} A_{nm} = -\frac{k_B T}{\hbar^2}(M\gamma_n\gamma_m + m_n\delta_{nm}), \tag{D31}$$

where $\gamma_0 = 1$, $m_0 = 0$. Finally, $p$-$p$ correlation is obtained

$$\langle \Delta p_n \Delta p_m \rangle = -\hbar^2 \left\langle \left( \frac{\partial}{\partial x_n} - \gamma_n \frac{\partial}{\partial Q} \right)\left( \frac{\partial}{\partial x_m} - \gamma_m \frac{\partial}{\partial Q} \right) \right\rangle$$

$$= k_B T(A_{nm} - \gamma_m A_{n0} - \gamma_n A_{m0} + \gamma_n\gamma_m A_{00})$$

$$= k_B T(M\gamma_n\gamma_m + m_n\delta_{mn} - M\gamma_m\gamma_n - M\gamma_n\gamma_m$$

$$+ M\gamma_n\gamma_m) = m_n k_B T \delta_{mn}. \tag{D32}$$

The above three results of $x$-$x$, $x$-$p$, and $p$-$p$ correlations are the same as Eq. (31).

### 3. Sufficient condition for "high" temperature

We assumed the high-temperature approximation $\frac{k_B T}{\hbar} \gg \frac{\sqrt{|\lambda_n|}}{2}$. Indeed, the temperature should satisfy $\frac{k_B T}{\hbar} \gg \frac{\sqrt{\lambda_M}}{2}$, where $\lambda_M$ is the absolute value of maximum eigenvalue of $A^{-1}B$. It is known that, for eigenvalue $\lambda$ of a matrix $A$, $|\lambda|$ is not greater than maximum column (or row) sum,[55]

$$|\lambda| \leq \max_j \sum_i |a_{ij}| \equiv \|A\|. \tag{D33}$$

According to the above definition of $\|\cdot\|$, It is not hard to see that $\|AB\| \leq \|A\|\|B\|$.

The above argument says

$$\lambda_M \leq \|A^{-1}B\| \leq \|A^{-1}\|\|B\|. \tag{D34}$$

It is not hard to obtain $A^{-1}$ with the following LDU factorization.

$$\begin{pmatrix} M & M\gamma_1 & M\gamma_2 & \cdots \\ M\gamma_1 & M\gamma_1^2 + m_1 & M\gamma_1\gamma_2 & \cdots \\ M\gamma_2 & M\gamma_2\gamma_1 & M\gamma_2^2 + m_2 & \cdots \\ \vdots & \vdots & \vdots & \ddots \end{pmatrix} = \begin{pmatrix} 1 & 0 & 0 & \cdots \\ \gamma_1 & 1 & 0 & \cdots \\ \gamma_2 & 0 & 1 & \cdots \\ \vdots & \vdots & \vdots & \ddots \end{pmatrix} \begin{pmatrix} M & 0 & 0 & \cdots \\ 0 & m_1 & 0 & \cdots \\ 0 & 0 & m_2 & \cdots \\ \vdots & \vdots & \vdots & \ddots \end{pmatrix} \begin{pmatrix} 1 & \gamma_1 & \gamma_2 & \cdots \\ 0 & 1 & 0 & \cdots \\ 0 & 0 & 1 & \cdots \\ \vdots & \vdots & \vdots & \ddots \end{pmatrix}. \tag{D35}$$

Inverting the factorized matrices,

$$A^{-1} = \begin{pmatrix} 1 & \gamma_1 & \gamma_2 & \cdots \\ 0 & 1 & 0 & \cdots \\ 0 & 0 & 1 & \cdots \\ \vdots & \vdots & \vdots & \ddots \end{pmatrix}^{-1} \begin{pmatrix} M & 0 & 0 & \cdots \\ 0 & m_1 & 0 & \cdots \\ 0 & 0 & m_2 & \cdots \\ \vdots & \vdots & \vdots & \ddots \end{pmatrix}^{-1} \begin{pmatrix} 1 & 0 & 0 & \cdots \\ \gamma_1 & 1 & 0 & \cdots \\ \gamma_2 & 0 & 1 & \cdots \\ \vdots & \vdots & \vdots & \ddots \end{pmatrix}^{-1} = \begin{pmatrix} 1 & -\gamma_1 & -\gamma_2 & \cdots \\ 0 & 1 & 0 & \cdots \\ 0 & 0 & 1 & \cdots \\ \vdots & \vdots & \vdots & \ddots \end{pmatrix} \begin{pmatrix} \frac{1}{M} & 0 & 0 & \cdots \\ 0 & \frac{1}{m_1} & 0 & \cdots \\ 0 & 0 & \frac{1}{m_2} & \cdots \\ \vdots & \vdots & \vdots & \ddots \end{pmatrix}$$

$$\times \begin{pmatrix} 1 & 0 & 0 & \cdots \\ -\gamma_1 & 1 & 0 & \cdots \\ -\gamma_2 & 0 & 1 & \cdots \\ \vdots & \vdots & \vdots & \ddots \end{pmatrix} = \begin{pmatrix} \frac{1}{M} + \sum_n \frac{\gamma_n^2}{m_n} & -\frac{\gamma_1}{m_1} & -\frac{\gamma_2}{m_2} & \cdots \\ -\frac{\gamma_1}{m_1} & \frac{1}{m_1} & 0 & \cdots \\ -\frac{\gamma_2}{m_2} & 0 & \frac{1}{m_2} & \cdots \\ \vdots & \vdots & \vdots & \ddots \end{pmatrix}. \tag{D36}$$

Thus, the maximum column sum of $A^{-1}$ is

$$\|A^{-1}\| = \max_n \left( \frac{1}{M} + \sum_i \frac{\gamma_i^2}{m_i} + \sum_i \frac{|\gamma_i|}{m_i}, \frac{1+|\gamma_n|}{m_n} \right). \quad (D37)$$

If $\gamma_i$ are on the order of 1 or larger, $\frac{1}{M} + \sum_i \frac{\gamma_i^2}{m_i} + \sum_i \frac{|\gamma_i|}{m_i}$ is the maximum value. And, in this limit, it is smaller than $\frac{1}{M} + 2\sum_i \frac{\gamma_i^2}{m_i}$. So one can get

$$\|A\| \leq \frac{1}{M} + 2\sum_i \frac{\gamma_i^2}{m_i}. \quad (D38)$$

Since $B$ is given by

$$B = \begin{pmatrix} b + \sum_n m_n \omega_n^2 & -m_1\omega_1^2 & -m_2\omega_2^2 & \cdots \\ -m_1\omega_1^2 & m_1\omega_1^2 & 0 & \cdots \\ -m_2\omega_2^2 & 0 & m_2\omega_2^2 & \cdots \\ \vdots & \vdots & \vdots & \ddots \end{pmatrix}, \quad (D39)$$

the maximum column sum of $B$ is

$$\|B\|_1 = \max_n \left( b + 2\sum_i m_i w_i^2, 2m_n w_n^2 \right) \leq |b| + 2\sum_i m_i w_i^2. \quad (D40)$$

Finally, one obtains the upper bound of $\lambda_M$,

$$\lambda_M \leq \|A^{-1}\|\|B\| \leq \left( \frac{1}{M} + 2\sum_i \frac{\gamma_i^2}{m_i} \right)\left( |b| + 2\sum_i m_i \omega_i^2 \right). \quad (D41)$$

In order to evaluate the expression on the right-hand side of the inequality Eq. (D41), we use the constraints Eq. (20). To convert the summations to known quantities, we generalize the constraint to the Caldeira-Legget-type continuous form with the following definitions of spectral functions,

$$J_p(\omega) \equiv \frac{\pi}{2}\sum_i \frac{\gamma_i^2 \omega_i}{m_i}\delta(\omega_i - \omega) = \frac{\alpha S}{2KM}\omega, \quad (D42)$$

$$J_x(\omega) \equiv \frac{\pi}{2}\sum_i m_i \omega_i^3 \delta(\omega_i - \omega) = \frac{2\alpha KM}{S}\omega. \quad (D43)$$

Checking the constraints,

$$\sum_i \frac{\gamma_i^2 \lambda}{m_i(\lambda^2 + \omega_i^2)} = \frac{2\lambda}{\pi}\int d\omega \frac{J_p(\omega)}{\omega(\lambda^2 + \omega^2)}$$
$$= \frac{2\lambda}{\pi}\frac{\alpha S}{2KM}\int d\omega \frac{1}{\lambda^2 + \omega^2} = \frac{\alpha S}{2KM}, \quad (D44)$$

$$\sum_i \frac{m_i \omega_i^2 \lambda}{\lambda^2 + \omega_i^2} = \frac{2\lambda}{\pi}\int d\omega \frac{J_x(\omega)}{\omega(\lambda^2 + \omega^2)}$$
$$= \frac{2\lambda}{\pi}\frac{2\alpha KM}{S}\int d\omega \frac{1}{\lambda^2 + \omega^2} = \frac{2\alpha KM}{S}. \quad (D45)$$

Finally,

$$\sum_i \frac{\gamma_i^2}{m_i} = \frac{2}{\pi}\int d\omega \frac{\alpha S}{2KM} = \frac{\alpha S}{\pi KM}\omega_c, \quad (D46)$$

$$\sum_i m_i \omega_i^2 = \frac{2}{\pi}\int d\omega \frac{2\alpha KM}{S} = \frac{4\alpha KM}{S\pi}\omega_c, \quad (D47)$$

where $\omega_c$ is the critical frequency of the environmental excitations.

Therefore, $\lambda_M \leq (\frac{1}{M} + \frac{2\alpha S}{\pi KM\omega_c})(|b| + \frac{8\alpha KM}{S\pi}\omega_c)$. Hence, one finally finds that the sufficient condition of the high temperature is $T \gg T_c$, where the critical temperature $T_c$ is defined as

$$T_c \equiv \frac{\hbar}{2k_B}\sqrt{\left(\frac{1}{M} + \frac{2\alpha S}{\pi KM}\omega_c\right)\left(|b| + \frac{8\alpha KM}{S\pi}\omega_c\right)}. \quad (D48)$$

Now, we check if the above condition is satisfied in experimental situations. Ignoring $|b|$, the critical temperature becomes $\sqrt{(1 + \frac{2\alpha S}{\pi K}\omega_c)\frac{2\alpha K}{S\pi}\omega_c}$. Since the environmental excitation is caused by magnetization dynamics, one can note that there is no need to consider the environmental excitation with frequencies far exceeding the frequency scale of magnetization dynamics. This concludes that $\omega_c$ is on the order of the frequency of magnetization dynamics, which is known as about 10 GHz or less.[56,57] With conventional scale $\alpha \sim 0.01$, $K \sim 10^{-4}$ eV, and $2S \sim \hbar$, the critical temperature is estimated as $T_c \sim 30$ mK. Therefore, our calculation is concluded to be well satisfied in most experimental situation.

---

KYOUNG-WHAN KIM AND HYUN-WOO LEE                                    PHYSICAL REVIEW B **82**, 134431 (2010)K.-H. Shin, S.-B. Choe, and H.-W. Lee, Nature (London) **458**, 740 (2009).

[9] R. A. Duine, A. S. Núñez, and A. H. MacDonald, Phys. Rev. Lett. **98**, 056605 (2007).

[10] R. A. Duine and C. M. Smith, Phys. Rev. B **77**, 094434 (2008).

[11] J.-V. Kim and C. Burrowes, Phys. Rev. B **80**, 214424 (2009).

[12] E. Martinez, L. Lopez-Diaz, L. Torres, C. Tristan, and O. Alejos, Phys. Rev. B **75**, 174409 (2007).

[13] E. Rossi, O. G. Heinonen, and A. H. MacDonald, Phys. Rev. B **72**, 174412 (2005).

[14] E. Martinez, L. Lopez-Diaz, O. Alejos, L. Torres, and C. Tristan, Phys. Rev. Lett. **98**, 267202 (2007).

[15] S.-M. Seo, K.-J. Lee, W. Kim, and T.-D. Lee, Appl. Phys. Lett. **90**, 252508 (2007).

[16] D. Ravelosona, D. Lacour, J. A. Katine, B. D. Terris, and C. Chappert, Phys. Rev. Lett. **95**, 117203 (2005).

[17] M. Hayashi, L. Thomas, C. Rettner, R. Moriya, X. Jiang, and S. S. P. Parkin, Phys. Rev. Lett. **97**, 207205 (2006).

[18] R. A. Duine, A. S. Núñez, J. Sinova, and A. H. MacDonald, Phys. Rev. B **75**, 214420 (2007).

[19] R. K. Pathria, *Statistical Mechanics* (Elsevier, Oxford, 1972).

[20] S. E. Barnes and S. Maekawa, Phys. Rev. Lett. **95**, 107204 (2005).

[21] H. Kohno, G. Tatara, and J. Shibata, J. Phys. Soc. Jpn. **75**, 113706 (2006).

[22] Y. Tserkovnyak, H. J. Skadsem, A. Brataas, and G. E. W. Bauer, Phys. Rev. B **74**, 144405 (2006).

[23] G. Tatara and H. Kohno, Phys. Rev. Lett. **92**, 086601 (2004).

[24] G. Tatara, H. Kohno, J. Shibata, Y. Lemaho, and K.-J. Lee, J. Phys. Soc. Jpn. **76**, 054707 (2007).

[25] G. Tatara, H. Kohno, and J. Shibata, J. Phys. Soc. Jpn. **77**, 031003 (2008).

[26] X. Waintal and M. Viret, Europhys. Lett. **65**, 427 (2004).

[27] A. Vanhaverbeke and M. Viret, Phys. Rev. B **75**, 024411 (2007).

[28] M. Thorwart and R. Egger, Phys. Rev. B **76**, 214418 (2007).

[29] I. Garate, K. Gilmore, M. D. Stiles, and A. H. MacDonald, Phys. Rev. B **79**, 104416 (2009).

[30] R. P. Feynman and F. L. Vernon, Jr., Ann. Phys. (N.Y.) **24**, 118 (1963).

[31] C. M. Smith and A. O. Caldeira, Phys. Rev. A **36**, 3509 (1987).

[32] A. O. Caldeira and A. J. Leggett, Phys. Rev. Lett. **46**, 211 (1981).

[33] G. L. Ingold and Yu. V. Nazarov, in *Single Charge Tunneling Coulomb Blockade Phenomena in Nanostructures*, edited by H. Grabert and M. Devoret (Plenum, New York, 1992).

[34] H. Lee and L. S. Levitov, Phys. Rev. B **53**, 7383 (1996).

[35] N. L. Schryer and L. R. Walker, J. Appl. Phys. **45**, 5406 (1974).

[36] A. P. Malozemoff and J. C. Slonczewski, *Magnetic Domains Walls in Bubble Materials* (Academic, New York, 1979).

[37] S.-W. Jung, W. Kim, T.-D. Lee, K.-J. Lee, and H.-W. Lee, Appl. Phys. Lett. **92**, 202508 (2008); J. Ryu and H.-W. Lee, J. Appl. Phys. **105**, 093929 (2009).

[38] M. Kläui, P.-O. Jubert, R. Allenspach, A. Bischof, J. A. C. Bland, G. Faini, U. Rüdiger, C. A. F. Vaz, L. Vila, and C. Vouille, Phys. Rev. Lett. **95**, 026601 (2005).

[39] G. S. D. Beach, C. Knutson, C. Nistor, M. Tsoi, and J. L. Erskine, Phys. Rev. Lett. **97**, 057203 (2006).

[40] M. Hayashi, L. Thomas, C. Rettner, R. Moriya, Y. B. Bazaliy, and S. S. P. Parkin, Phys. Rev. Lett. **98**, 037204 (2007).

[41] For permalloy, $|e\gamma_0 K\lambda/\mu_B| \sim 10^9$ A/cm$^2$, which is about an order larger than the current density of $\sim 10^8$ A/cm$^2$ used in many experiments (Refs. 38–40).

[42] Y. Le Maho, J.-V. Kim, and G. Tatara, Phys. Rev. B **79**, 174404 (2009).

[43] T. Kim, J. Ieda, and S. Maekawa, arXiv:0901.3066 (unpublished).

[44] V. W. Döring, Z. Naturforsch. A **3A**, 373 (1948).

[45] We thank M. Stiles for pointing out this point.

[46] To solve this system, one of the constraints Eq. (20a) is generalized to hold even for $\lambda=0$. That is, $\Sigma_i \gamma_i = 0$. See, for a detail, Appendix C.

[47] To consider a force on Eq. (6a), the potential should be generalized to depend on the momentum.

[48] For $v=0$, the terminal velocity of the DW vanishes independently of its the initial velocity since the environmental mass is much larger than the DW mass. With $v>0$, one can perform the Galilean transformation to make $\langle \dot{x}_i(0)\rangle=0$ instead of $\langle \dot{x}_i(0)\rangle=v$. Since the system is Galilean invariant, one expect that the DW also stops in this frame, just as $v=0$. It implies that the terminal velocity of the DW in the lab frame is also $v$.

[49] V. Kamberský, Czech. J. Phys., Sect. B **26**, 1366 (1976); Can. J. Phys. **48**, 2906 (1970); Czech. J. Phys., Sect. B **34**, 1111 (1984).

[50] K. Gilmore, Y. U. Idzerda, and M. D. Stiles, Phys. Rev. Lett. **99**, 027204 (2007).

[51] C.-Y. You, I. M. Sung, and B.-K. Joe, Appl. Phys. Lett. **89**, 222513 (2006); C.-Y. You and S.-S. Ha, *ibid.* **91**, 022507 (2007).

[52] This section summarizes the work by Kim *et al.* (Ref. 43).

[53] J. M. Winter, Phys. Rev. **124**, 452 (1961).

[54] By the same argument, Eq. (32) is obtained under an arbitrary potential $V(Q)$. Since the system was in equilibrium before applying current, we assume $V'(Q)=0$. At high temperature limit, $\chi$ moves in very short (imaginary) time interval. Therefore, we can take quadratic approximation and $V(Q)$ to be the form of $bQ^2$.

[55] See, for example, G. Strang, *Linear Algebra and its Applications* (Thomson, USA, 1988), Chap. 7.

[56] A. Mourachkine, O. V. Yazyev, C. Ducati, and J.-Ph. Ansermet, Nano Lett. **8**, 3683 (2008).

[57] C. Boone, J. A. Katine, J. R. Childress, J. Zhu, X. Cheng, and I. N. Krivorotov, Phys. Rev. B **79**, 140404(R) (2009).

134431-16